%% file: main.tex
\documentclass[aps,prl,twocolumn,preprintnumbers,superscriptaddress]{revtex4-1}

\usepackage{amsmath, mathrsfs, amssymb,amsfonts,amsthm,graphicx, epsf, dcolumn, yfonts}
\usepackage[normalem]{ulem}
\usepackage[hyperfootnotes=true]{hyperref}
\usepackage{color}
\usepackage{amssymb}
\usepackage{slashed}
\usepackage{setspace}
\usepackage{cancel}
\usepackage{wasysym}
\usepackage{float}
\usepackage[utf8]{inputenc}
\usepackage{todonotes}

        \hypersetup{
           breaklinks=true,   
           colorlinks=true,   
        }

\pdfoutput=1
\newcommand{\lin}{L}

\input{definitions}

\def\z{{ q}}

\begin{document}

\title{Quantum gravity with dynamical wave-function collapse via a classical scalar field}

\author{Zachary Weller-Davies}
\affiliation{InstaDeep, London, W2 1AY, United Kingdom}

\begin{abstract}
\noindent In hybrid classical-quantum theories, the dynamics of the classical system induce the classicality of the quantum system, meaning that such models do not necessarily require a measurement postulate to describe probabilistic measurement outcomes. It has recently been shown that covariant classical-quantum dynamics can be constructed using path integral methods, with the dynamics encoded in a combined action for the classical and quantum variables. This work introduces a classical-quantum model whereby quantum gravity interacts with a classical scalar field. The scalar field can be viewed as fundamentally classical or effectively classical due to the decoherence of a quantum gravity theory. The dynamics act to collapse quantum spacetimes according to their Ricci scalar, with corresponding diffusion in the scalar field due to the quantum back-reaction. In the classical limit, the diffusion in the scalar field manifests itself via stochastic fluctuations in the Newtonian potential. Because we couple a classical scalar field to perturbative quantum gravity, we find the theory is not renormalizable but is instead to be viewed as an effective field theory. However, it is an effective field theory that does not necessarily require a measurement postulate. More generally, our work shows it is possible to integrate collapse dynamics with high-energy physics and covariance.

\end{abstract}

\maketitle

\section{Introduction} The tension between the unitary evolution of quantum systems and the non-unitary dynamics of measurement has still not been resolved precisely \cite{Adlam_2023, wallace2011decoherence, Frauchiger_2018}. In the search for a complete theory, unitary dynamics on the quantum state is often assumed from the start. However, in practice, the effective dynamics of experiments are not described unitarily but by probabilities according to the Born rule. To make experimental predictions in unitary quantum theory, one is seemingly forced to make a classical-quantum split, even though the dynamics of the quantum system and measurement device can, in principle, be described via quantum Hamiltonian's.

In making a classical-quantum split, usually, it is argued that effectively classical systems emerge from unitary quantum state evolution via considerations of decoherence with an environment \cite{wallace2011decoherence,sep-qm-manyworlds,wallace2009formal,carroll2018maddog,Rovelli_1996,Zurek_2003}. However, there is still the question of describing why non-unitary probabilistic dynamics occurs when effectively classical measurement devices are coupled to quantum systems, but not when isolated quantum systems are considered \cite{Adlam_2023, wallace2011decoherence, Frauchiger_2018}. Collapse models \cite{Pearle:1988uh,PhysRevA.42.78,BassiCollapse,PhysRevD.34.470, GisinCollapse, relcollapsePearle, 2006RelCollTum} consider the alternative viewpoint that the dynamics of quantum systems are not unitary but that the unitary evolution of the wave function describing the state of a quantum system is approximate. In collapse models, the quantum state evolution is supplemented with additional non-linear and stochastic terms that reproduce the probabilities of the Born rule. In both measurement and collapse models, though the evolution of the quantum state is non-linear, when averaged over the stochasticity, the density matrix evolution is linear, and probabilities are conserved. To have consistency with experiment, an amplification mechanism is required to ensure that the decoherence rate is more pronounced in macroscopic systems. In contrast, the dynamics should remain approximately unitary for small isolated systems.

 Historically, collapse theories that aim to explain the disparity between unitary quantum dynamics and non-unitary measurement dynamics have been criticized because they lack a description in terms of a fundamental theory and are typically not presented in a covariant form \cite{Adlam_2023,wallace2022sky,Di_si_2022}. Though attempts have been made \cite{relcollapsePearle, 2006RelCollTum}, the lack of integration of collapse dynamics with the principles of quantum field theory has, at large, spurred researchers to consider unitary fundamental theories. Probabilities are then argued to emerge from unitary quantum state evolution \cite{wallace2011decoherence,sep-qm-manyworlds,wallace2009formal,carroll2018maddog,Rovelli_1996,Zurek_2003}. {Non-unitary open quantum theories have also been considered in the context of the black-hole information paradox \cite{bps, Srednicki_1993, Unruh_2017, Hawking_1976, Almheiri_2013} but these are out of favour since they violate conservation of the stress-energy tensor \cite{bps}.

Recently, covariant theories of classical fields coupled to quantum fields have been introduced using path integral methods \cite{Oppenheim:2023izn}. The classical-quantum (CQ) path integral is a generalization of the Feynman path integral for quantum systems \cite{FeynmanVernon}, and the stochastic path integral used to study classical stochastic processes \cite{Weber_2017, Kleinert}, and allows for interaction between the classical and quantum systems \cite{UCLPILONG, Oppenheim:2023izn}. In CQ theories, the interaction of the classical field with the quantum one induces classicality in the quantum state \cite{Oppenheim:2018igd, dec_Vs_diff, oppenheim2020objective, UCLHealing}. Such models can be viewed as effective, whereby an effectively classical system interacts with a quantum one within the limit of a fully quantum system. Alternatively, if the classical system is taken to be fundamental, then CQ theories provide dynamical realizations of collapse models, whereby the dynamics of the classical system induce the classicality of the quantum system \cite{Oppenheim:2018igd,oppenheim2020objective, UCLHealing, BLANCHARD1993272}.

CQ dynamics is described by completely positive evolution on combined classical-quantum states and yields positive and normalized probabilities \cite{poulinKITP2, Oppenheim:2018igd, blanchard1995event, diosi1995quantum}. The evolution is non-unitary, but in the same way that standard quantum theory is non-unitary when measurements are introduced. Furthermore, when the decoherence-diffusion trade-off \cite{dec_Vs_diff} is saturated, the dynamics maintain the purity of the quantum states when conditioned on the value of the classical field \cite{UCLHealing}. In particular, though the dynamics are stochastic, there is no loss of quantum information conditioned on the classical variable \cite{UCLHealing}. This is most evident in the continuous unraveling approach to CQ dynamics \cite{UCLHealing}, where it can be shown that the dynamics are formally equivalent to those governing the evolution of continuous quantum measurement and classical feedback \cite{2006Jacob, wiseman_milburn_2009}. However, the CQ equations can have a very different physical interpretation from continuous measurement: the classical degrees of freedom can be any classical system, and they can be dynamic, for example, the classical position and momenta of a classical system.

Covariant classical-quantum path integrals were introduced in order to construct theories of classical gravity interacting with quantum matter \cite{Diosi2014, 2016Tilloy,Oppenheim:2018igd, layton2023weak, pqconstraints}. In particular, \cite{Oppenheim:2023izn} studied a theory of classical gravity coupled to quantum matter that reproduces the trace of Einstein's equation in the classical limit. For the CQ interaction to be consistent with the stress-energy tensor sourcing the gravitational field, the theory predicts that superpositions decohere at a rate dependent on the difference in stress-energy tensors of the superposed states. This ensures that coherence can be potentially maintained for small masses but not large macroscopic objects. Nonetheless, though much progress has been made, the theory in \cite{Oppenheim:2023izn} is not yet a complete theory consistent with observations since it is not known how to get all the components of the Einstein equation whilst also preserving the complete positivity of the dynamics.

 In this work, we come from an alternate direction. Motivated by constructing a dynamical collapse model, we consider perturbative \textit{quantum gravity} coupled to a classical scalar field. For simplicity, in this work, we consider a non-minimal Brans-Dicke coupling \cite{Brans_1961}. We let the classical scalar field and quantum metric interact dynamically so that there is both an effect of the classical scalar field on the quantum metric and a back-reaction of the quantum metric on the classical scalar field. To have such a back-reaction and maintain complete positivity, the dynamics are necessarily non-unitary and instead described by completely positive CQ dynamics which we review in Section \ref{sec: CQdynamics}.
 
 The path integral in Equation \eqref{eq: CQBD} describes the combined classical-quantum dynamics, and its weak field limit is described in Equation \eqref{eq: fullWF}. Unitarity is only approximate when quantum spacetimes with small Ricci curvature are considered. In particular, the classical scalar field dynamically induces the collapse of the wavefunction according to the Ricci scalar of the quantum spacetimes. This is an amplification mechanism by which macroscopic objects - associated with a larger Ricci scalar - decohere quickly, while quantum spacetimes with a small Ricci scalar can maintain coherence. There is corresponding randomness in the scalar field due to the quantum back-reaction. In the classical limit, the diffusion in the scalar field manifests itself via stochastic fluctuations in the Newtonian potential. We explore some of the low energy predictions of the theory in Section \ref{sec: pred}.

The theory is described by a diffeomorphism invariant action and can be integrated with current high-energy physics since the matter sector of the theory is unitary. However, we find that it is likely not renormalizable, and so it should be viewed as an effective theory in a similar way to perturbative quantum gravity \cite{Veltman:1975vx,tHooft:1974toh,Donoghue_1994,Burgess:2003jk}. Here, by effective theory, we mean that the quantum gravity sector of the theory is to be viewed as an effective theory, not that the classical scalar field is necessarily described as an effectively classical limit of a fully quantum theory: the UV complete theory could be either fully quantum or CQ. Though the CQ path integral takes the form of a type of quantum field theory, effective CQ theories have not been studied in the literature. To gain some intuition, in Appendix \ref{sec: effective}, we consider a simple example of a top-down effective CQ theory. There, we start from a CQ path integral consisting of a classical degree of freedom and two quantum degrees of freedom that all interact with each other. We then integrate out one of the quantum systems and show the resulting effective action is described by a CQ path integral, but with the addition of an infinite series of terms that are suppressed by the mass scale of the particle that has been integrated out. We therefore expect that one can treat effective CQ theories on a similar footing to standard effective quantum field theories \cite{GeorgiEFT}.

In this work, we make the specific choice of non-minimal Brans-Dicke \cite{Brans_1961} coupling since it is calculationally tractable. A general class of theories that we expect to have similar features to that considered here would be based on the Horndeski theory \cite{Horndeski1974}. Hordenski's theory is the most general theory of scalar-tensor gravity that leads to second-order equations of motion. It contains Brans-Dicke \cite{Brans_1961} and many other theories often considered in cosmological models \cite{Kobayashi_2019}. Since Hordenski theories have second-order equations of motion for both the scalar field and metric tensor, we expect that, like the theory considered here, classical-quantum path integrals (i.e., Equation \eqref{eq: CQProto}) will also lead to decoherence rates that depend on the curvature of the quantum spacetime. It would be interesting to explore the low energy limit of such CQ theories and their implications for cosmology \cite{Clifton_2012,Peracaula_2020}.

We interpret our results as a proof of principle that collapse models can be made consistent with notions of quantum field theory. In particular, we find that the low-energy physics of quantum measurement can be explained through a dynamical classical scalar field that interacts with quantum gravity gravity. This is an example of a model whereby quantum gravity plays a role in wavefunction collapse \cite{Penrose:1996cv,Disi1987} -- via its interaction with a classical field. Further studying such theories' dynamics, naturalness in the context of UV complete quantum gravity, and how to treat them in the framework of effective field theories, is crucial in understanding their validity. This is beyond the scope of the current work; the goal of the present article is to construct a quantum gravity model with dynamical collapse and to motivate further studies of such theories.

 The paper's outline is as follows: In Section \ref{sec: CQdynamics}, we introduce the CQ path integral formalism used throughout the paper. We refer to \cite{UCLPILONG, Oppenheim:2023izn} for the reader unfamiliar with CQ path integrals, but we also include extra details on what the CQ path integral is computing and how to ensure it is normalized. In Section \ref{sec: Born}, we show how measurement probabilities arise from the dynamics of CQ path integrals. In Section \ref{sec: QGScalar}, using the CQ path integral, we couple quantum gravity to a classical scalar field via a non-minimal Brans-Dicke-like coupling. We show that this leads to an amplification mechanism by which quantum states with larger Ricci scalar decohere, and we find an associated diffusion in the effective Newtonian constant. In Section \ref{sec: pred}, we consider the perturbative weak field limit and use it to summarize some of the theory's low energy predictions. We see that the coupling of the metric to the scalar field induces higher-order couplings to the Ricci scalar. In Section \ref{sec: renormalization}, we study the consequences of the CQ coupling on the renormalization and stability properties of the theory. We argue that we should view the theory as an effective theory, where the CQ action is the leading order term in an effective action. We conclude by discussing our results in Section \ref{sec: discussion}.
 
 In Appendix \ref{section: normalizing}, we discuss the normalization of the CQ path integral. In Appendix \ref{sec: effective}, we consider an example of a top-down approach to arrive at an effective CQ theory. There, we consider a classical system interacting with two quantum systems and study the effective action of the CQ system when one of the quantum systems is integrated out.

\section{Classical-quantum path integrals}\label{sec: CQdynamics} In this section, we introduce the formalism used to describe the consistent interaction of classical and quantum degrees of freedom \cite{Oppenheim:2018igd, poulinKITP2}. We first introduce the general formalism used to describe a classical degree of freedom coupled to a quantum one, and we denote a generic classical degree of freedom by $\q$; for example, the configuration of a classical particle or a scalar field. We also include detailed discussion on what the path integral computes, and how to recover measurement-like dynamics from the CQ path integral, which has previously been discussed via CQ unravelings \cite{UCLHealing, oppenheim2020objective, BLANCHARD1993272}. The reader familiar with CQ path integrals can skip to Section \ref{sec: QGScalar}.

When considering a hybrid system, the natural states are hybrid classical-quantum (CQ) states. Formally, a classical-quantum state associates to each classical variable an un-normalized density matrix $\cqstate(q,t) =p(q,t) \sigma(q,t)$ such that $\Tr_{\mathcal{H}}{\cqstate(q)} = p(q,t) \geq 0$ is a normalized probability distribution over the classical degrees of freedom and $\int d\q \cqstate(q,t) $ is a normalized density operator on a Hilbert space $\mathcal{H}$. Intuitively, $p(\z,t)$ can be understood as the probability density of being in the phase space point $\z$ and  $\sigma(\z,t)$ as the normalized quantum state one would have given the classical state $\z$ occurs.

Classical-quantum dynamics can then be understood as the set of linear dynamics which maps CQ states to CQ states. Linearity is required to maintain a probabilistic interpretation of the density matrix. The dynamics must be completely positive since we require that CQ  states be mapped to CQ states even when the dynamics act on half an entangled quantum state. In analogy with Krauss theorem for quantum operations, the most general form of CP dynamics mapping CQ states onto themselves is described by \cite{Oppenheim:2018igd, UCLPawula} \begin{equation}\label{eq: CPmap}
 \cqstate(\z,t+ \delta t) = \int d\z' \Lambda^{\mu\nu}(\z|\z',\delta t) \lin_{\mu}\cqstate(\z',t) \lin_{\nu}^{\dag},
\end{equation}  
where the $ \Lambda^{\mu \nu}(\z|\z',\delta t)$ defines a positive matrix-measure in $q,q'$. In Equation \eqref{eq: CPmap}, the operators $\lin_{\mu}$ are an arbitrary set of operators on the Hilbert space, and normalization of probabilities requires
\begin{equation}\label{eq: prob}
\int d\z  \Lambda^{\mu\nu}(\z|\z',\delta t) \lin_{\nu}^{\dag}\lin_{\mu} =\mathbb{I}.
\end{equation}

When the dynamics are time-local and autonomous, completely positive CQ master equations can be derived from Equation \eqref{eq: CPmap} and have been studied in \cite{alicki2003completely, poulinKITP2, Oppenheim:2018igd,oppenheim2020objective, Diosi:2011vu, Oppenheim:2018igd,pqconstraints,dec_Vs_diff,Di_si_2023}.

 Writing the dynamics in a path integral form is useful for studying CQ theories in a relativistic context. For a further discussion on CQ path integrals, we refer the reader to \cite{UCLPILONG, Oppenheim:2023izn}, while for the reader interested in further understanding CQ models and their relevance to collapse models, we refer the reader to \cite{UCLHealing, oppenheim2020objective, DiosiMeasurement}. In particular, \cite{UCLHealing} gives a description of CQ dynamics is given in terms of pure quantum states and stochastic classical trajectories. 

The path integral tells us how the components of the CQ state evolve. We denote quantum systems by $\psi$ and classical variables by $q$. In Section \ref{sec: QGScalar}, $\psi$ will become a quantum metric $g_{\mu\nu}$ and $q$ a classical scalar field $\phi$. 

Including explicitly a classical dynamical variable $q$, the path integral tells us how to evolve the components of the CQ state 
\begin{equation}\label{eq: statecomponents}
    \cqstate( q) = \int d \phi^+ d\psi^- \cqstate( q,\psi^+,\psi^-) \ | \psi^+ \rangle \langle \psi^- |,
\end{equation}
where $\psi^+$ and $\psi^-$ are the variables associated with the bra and ket components of the density matrix in a doubled integral. For the reader unfamiliar with density matrix path integrals, we recommend \cite{Sieberer_2016, Haehl_2017, Baidya:2017eho}. 

With Equation \eqref{eq: statecomponents} in mind, generically, a classical-quantum path integral will take the form \cite{Oppenheim:2023izn, UCLPILONG}
\begin{equation}\label{eq: transition}
\begin{split}
        \rho(q,\psi^+_f,\psi^-_f,t_f)  & = \int^{B} \mathcal{D}q \mathcal{D} \psi^+ \mathcal{D} \psi^- e^{I[q,\psi^+,\psi^-,t_i,t_f]}   \\
        & \times \rho(q,\psi^+_i,\psi^-_i, t_i).
\end{split}
\end{equation}
In Equation \eqref{eq: transition}
\begin{equation}\label{eq: boundarycondition}
   B= \{q(t_f) =q_f, \psi^+(t_f) = \psi^+_f, \psi^-(t_f) = \psi^-_f\}
\end{equation} denotes the boundary conditions at $t_f$. Equation \eqref{eq: transition} describes transition amplitude to go from one CQ state to another as an integral over possible configurations of the classical variable $q$, and bra-ket variables $\psi^+, \psi^-$, weighted by the CQ action.

Classical quantum path integrals for completely positive master equations have recently been studied in \cite{Oppenheim:2023izn, UCLPILONG}. Of particular relevance is the result that any classical-quantum configuration space path integral with an action of the form 
 \begin{equation}\label{eq: positiveCQ}
 \begin{split}
  I( \q, \xqr,  \xql,t_i,t_f) &=  I_{CQ}(\q,\psi^+, t_i,t_f)\\
  & + I^*_{CQ}(\q,\psi^-, t_i,t_f) - I_C(\q, t_i,t_f)
  \end{split}
\end{equation}
 defines completely positive CQ dynamics \cite{Oppenheim:2023izn}. Ensuring the path integral is normalized is trickier. We discuss this in more detail in Section \ref{sec: protoreview} and in Appendix \ref{section: normalizing}. A general form of path integral which preserves the norm of quantum states is derived from master equation methods in \cite{UCLPILONG}. In Equation \eqref{eq: positiveCQ} $I_{CQ}^{\pm}$ determines the CQ interaction on each branch, and $I_C$ is a purely classical Fokker-Plank like action \cite{Kleinert, Sieberer_2016} which should be positive semi-definite. We can also allow for the case where some of the classical degrees of freedom are deterministic and not associated with back-reaction, in which case the purely classical part of the path integral contains additional delta functions. 

 For completeness, we give a simple argument as to why the path integral gives rise to positive probabilities through completely positive (CP) evolution \cite{Oppenheim:2023izn, UCLPILONG}. 
 
 Positivity of the CQ state means that for any Hilbert space vector $|v(q)\rangle $ we have $\Tr[|v(q)\rangle \langle v(q)| \cqstate(q)] \geq 0 $. This ensures the density matrix $\cqstate(q)$ is a positive matrix operator for each $q$, which implies that probabilities are positive. In components, positivity is equivalent to asking that for any vector $|v(q)\rangle$, with components $v(\psi, q) = \langle \psi|v(q) \rangle$, we have
\begin{equation}\label{eq: CPpath}
    \int d\psi^{+} d\psi^- v(\psi^+,q)^* \cqstate(\psi^+,\psi^-,\q) v(\psi^-,q) \geq 0 .
\end{equation}
A CQ dynamics $\Lambda$ is said to be positive if it preserves the positivity of CQ states and completely positive if $\mathbb{I} \otimes \Lambda$ is positive when we act with the identity on any larger system. 

We now show that each field integral in Equation \eqref{eq: positiveCQ} preserves the positivity of the quantum state for each field integral, which we label by $k$ \cite{Oppenheim:2023izn}. Inserting the path integral of Equation \eqref{eq: CQProto} into the definition of positivity in Equation \eqref{eq: CPpath}, we see the positivity of the state at step $k+1$ factorizes according to 
\begin{equation}\label{eq: transition2}
\begin{split}
    & \int  d\psi^{+}_{k+1} d\psi^-_{k+1} dq_k d\psi^+_k d\psi^-_k \\
    & \times [v(\psi^+_{k+1},q_{k+1})^* e^{ I_{CQ}^+}][v(\psi^-_{k+1},q_{k+1})^*e^{ I_{CQ}^-}]^* \\
&\times e^{ - I_{C}}\cqstate(\psi^+_k,\psi^-_k,\q_k,t_k).
\end{split}
\end{equation}
The two terms containing $e^{I_{CQ}}$ come in conjugate pairs, and so redefining $\bar{v} = v e^{I_{CQ}^*}$ we can make comparison with Equation \eqref{eq: CPpath}, noting that $e^{-I_C} \geq 0$ if $I_C$ is positive. From Equation \eqref{eq: transition2}, we see that if the state at step $k$ is positive, then the integral over step $k$ will yield a positive function over the $k+1$ variables. Hence, the state at step $k+1$ will be positive if the state at step $k$ is positive. Thus, if the initial state is positive, the final state from the path integral in Equation \eqref{eq: positiveCQ} will be positive. Complete positivity of the dynamics follows almost immediately, since when acting on part of an entangled system, the dynamics still factorize in this way \cite{Oppenheim:2023izn}. 

\subsection{What the path integral computes}
This section discusses how the CQ path integral can be used to compute correlation functions and transition amplitudes. 

\subsubsection{CQ joint states}
First, note that we can use the path integral to describe joint probability distributions over the classical variable. Classically, the joint probability distribution $p(q_2, t_2, q_1, t_1)$ describes the probability distribution of finding $q_1$ at $t_1$ and $q_2$ at $t_2$. When a quantum state interacts with the classical system, we will also have an associated normalized density matrix $\sigma(q_2, t_2, q_1, t_1)$. This density matrix describes the quantum density matrix, given we find the classical state in $q_1$ at $t_1$ and $q_2$ at $t_2$. The weight of this density matrix is given by the classical joint probability distribution so that the combined CQ state is then defined by the product 
\begin{equation}\label{eq: jointCQ}
  \cqstate(q_2, t_2, q_1, t_1)= p(q_2, t_2, q_1, t_1)\sigma(q_2, t_2, q_1, t_1).
\end{equation}
The components of the joint CQ state are described by the path integral
\begin{equation}\label{eq: cqjoint}
\begin{split}
& \cqstate(q_2,\psi^+_2, \psi^-_2,t_2, q_1, t_1) =\\
& \int^{q(t_2) = q_2}_{q(t_1) = q_1} \mathcal{D} q \mathcal{D}\psi^+ \mathcal{D} \psi^- e^{  \mathcal{I}( \q, \xqr,  \xql,t_1,t_2)}\\
& \times \int^{q(t_1) = q_1} \mathcal{D} q \mathcal{D}\psi^+ \mathcal{D} \psi^- e^{  \mathcal{I}( \q, \xqr,  \xql,t_i,t_1)} \cqstate(q_0, \psi^+_i, \psi^-_i, t_i).
\end{split}
\end{equation}
In Equation \eqref{eq: cqjoint}, we first perform the CQ path integral over all classical and quantum configurations, where the final classical state is $q_1$, and then we glue this to a path integral over all configurations where the final classical state is $q_2$; the $q_1,q_2$ boundary conditions are not integrated over.

Another way of understanding the joint distribution via the path integral approach is that we performed the path integral up to time $t_2$ and inserted delta functions $\delta(q(t_2) - q_2) \delta(q(t_1)-q_1)$. 
In this way, one can further define a joint distribution over the classical and quantum states by gluing a series of path integrals together and inserting any combinations of delta functions. For example, consider the case where we insert delta functions at specific times $t_n,\dots t_1$
\begin{equation}
\begin{split}
        & \delta_{n,\dots,1} =  \delta(q(t_{_n}) - q_n)\delta(\psi^+(t_n) - \psi^+_n) \delta(\psi^-(t_n) - \psi^-_n)\\
        & \dots \times  \delta(q(t_1) - q_1)\delta(\psi^+(t_1) - \psi^+_1) \delta(\psi^-(t_1) - \psi^-_1).
\end{split}
\end{equation}
In this case, the joint CQ state is defined by inserting $\delta_{n,\dots,1}$ into Equation \eqref{eq: transition}
\begin{equation}
\begin{split}
 &  \cqstate(q_n,\psi^+_n, \psi^-_n,t_n,\dots, q_1,\psi^+_1, \psi^-_1,t_1) = \\
 & \int \mathcal{D}q \mathcal{D} \psi^+ \mathcal{D} \psi^- e^{\mathcal{I} [q,\psi^+,\psi^-,t_i,t_n]}  \delta_{n,\dots,1}\rho(q,\psi^+_i,\psi^-_i, t_i).
\end{split}
\end{equation}
More generally, one can insert delta functions of any bra, ket, and classical variables at any time. Since the CQ state is normalized, and the joint CQ state is found by inserting delta functions, the joint CQ state is normalized. However, it is not, in general, a positive operator. The joint CQ state is a generalization of transition amplitudes in the combined classical and quantum case and computes the matrix elements of the CQ state. As such, they need not define positive operators. 

We can compute overlaps between joint CQ states by multiplying them together in conjugation, meaning that we take the conjugation and swap all bras to kets in one of the states and integrate over all the variables. As a simple example, we can compute the overlap between the two CQ states of the form in Equation \eqref{eq: cqjoint} through
\begin{equation}\label{eq: overlap}
\begin{split}
& \int dq_1 \dots dq_n dq'_1 \dots dq'_m  \\
& \Tr[\cqstate(q_n, t_n, \dots, q_1,t_1) \cqstate(q'_m, t'_m, \dots, q'_1,t'_1)  ].
\end{split}
\end{equation} 
When the classical distribution is a delta function, meaning that the joint distributions are represented by delta functions over the classical variables, Equation \eqref{eq: overlap} reduces to the normal quantum state overlap definition. 

Using the joint distribution, we can compute correlation functions of classical-quantum observables $O(q_n, t_n)$
\begin{equation}\label{eq: correlation}
\int dq_1 \dots dq_n \Tr[O(q_n,t_n) \dots O(q_1,t_1) \cqstate(q_n, t_n, \dots,.q_1,t_1) ],
\end{equation} 
where $O$ is a quantum operator. We will write these as $\langle O(q_n,t_n) \dots O(q_1,t_1)\rangle_{\cqstate_i}$.
When we compute the expectation value of such operators, we insert the operators into Equation \eqref{eq: cqjoint} and integrate 
over the classical variables. We also integrate over the final state. Correlators in Equation \eqref{eq: correlation} are thus computed by the CQ path integral 
\begin{equation}
\begin{split}
   &  \int^{\psi_f^+ = \psi_f^-} \mathcal{D}q \mathcal{D}\psi^+ \mathcal{D}\psi^-  e^{  \mathcal{I}( \q, \xqr,  \xql,t_i,t_f)}O(q_n,t_n) \dots O(q_1,t_1)\\
    & \times \cqstate(q_i, \psi^+_i, \psi^-_i, t_i),
    \end{split}
\end{equation}
where in the path integral, the Schwinger-Keldysh boundary conditions $\psi^+_f = \psi^-_f$ are imposed to enforce the trace of the time-evolved system. 
\subsubsection{Scattering amplitudes and partition functions}
We can also compute correlation functions conditioned on a final CQ state by imposing final time boundary conditions
\begin{equation}\label{eq: correlation2}
\begin{split}
& {}_{\cqstate_f}\langle O(q_n,t_n) \dots O(q_1,t_1)\rangle_{\cqstate_i} :=\int dq_1 \dots dq_n \\
& \Tr[\cqstate(q_f) O(q_n,t_n) \dots O(q_1,t_1) \cqstate(q_n, t_n, \dots, q_1,t_1) ].
\end{split}
\end{equation} 
Equation \eqref{eq: correlation2} is computed via the path integral 
\begin{equation}
\begin{split}
   &  \int^{B} \mathcal{D}q \mathcal{D}\psi^+ \mathcal{D}\psi^-  e^{  \mathcal{I}( \q, \xqr,  \xql,t_i,t_f)}O(q_n,t_n) \dots O(q_1,t_1)\\
    & \times \cqstate(q_i, \psi^+_i, \psi^-_i, t_i),
    \end{split}
\end{equation}
where $B$ are the boundary conditions on the final state defined in Equation \eqref{eq: boundarycondition}.

Correlation functions, such as those in Equation \eqref{eq: correlation}, can be found by inserting sources into the partition function of the CQ system. The partition function is computed by taking the trace over the final state 
\begin{equation}
\begin{split}    
& Z_{t_f,t_i}  = \int dq_f d \psi^+_f d \psi^-_f \delta(\psi^+_f-\psi^-_f) \cqstate(q_f,\psi^+_f, \psi^-_f, t_f) \\
& =\int^{\psi^+_f=\psi^-_f} \mathcal{D}q \mathcal{D}\psi^+ \mathcal{D}\psi^- e^{  \mathcal{I}( \q, \xqr,  \xql,t_i,t_f)} \cqstate(q,\psi^+_i, \psi^-_i, t_i),
\end{split}
\end{equation}
where in the path integral, the Schwinger-Keldysh boundary conditions $\psi^+_f = \psi^-_f$ are imposed to enforce the trace of the system. We define the partition with sources as 
\begin{equation}
    Z_{t_f,t_i}[ J_+, J_-, J_q] = \langle e^{i J_+ \psi^ - -  i J_- \psi^- - J_q q} \rangle_{\rho_i},
\end{equation}
in which case correlation functions can be computed using standard functional integral methods. 

If we restrict ourselves to starting in the stationary state of the system but want to consider correlation functions at arbitrarily long times, we take the limit $t_i\to -\infty, t_f \to \infty$. A typical assumption in an open system is that the initial state in the infinite past doesn't affect the stationary state, i.e., it is unique \cite{Sieberer_2016}. In this case, we can ignore the initial boundary state and obtain the partition function 
\begin{equation}
\begin{split}    
Z =\int^{\psi^+(\infty)=\psi^-(\infty)} \mathcal{D}q \mathcal{D}\psi^+ \mathcal{D}\psi^- e^{  \mathcal{I}( \q, \xqr,  \xql,t_i,t_f)}.
\end{split}
\end{equation}

However, the assumption of a unique steady state is strong and generally isn't the case when the CQ dynamics has competing terms. If a unique steady state doesn't exist, the next best thing one can consider is the subspace of asymptotically stable states, and we label an element of this subspace parametrically by $s$, with $\cqstate_s \in \mathcal{S}$. For any initial CQ state, we take the asymptotically stable subspace $\mathcal{S}$ to be the space of CQ states $\cqstate_s$ that do not have vanishing probability amplitude $\Tr[\cqstate_s(q)] \approx 0$ as we take the initial state evolution to $-\infty$. Elements in this subspace will generally still interact with each other.

Transition amplitudes $ {}_{\cqstate_s'}\langle 1 \rangle_{\cqstate_s}$ between asymptotically steady states are the physical things that we can imagine preparing in an experiment. In quantum field theory, we take these to be in-out eigenstates of a Hamiltonian. For the CQ theory they are very system-dependent. Strong measurement dynamics has a steady state on the quantum - the eigenstate of the measurement operator - but weak measurement dynamics have a much more complex system of steady states if the Hamiltonian term doesn't commute with the operator being measured. It should be said such states are not really steady since they interact with each other. We leave a detailed study of the CQ asymptotic steady states for future work. Many of these questions are perhaps best answered in the operator formalism, but we choose to work via path integrals to make covariance manifest. The relationship between CQ master equations and path integrals has been worked out in \cite{UCLPILONG}, and it would be interesting to study this in more detail for the theory presented here. 

\subsection{The proto action}\label{sec: protoreview}
In \cite{Oppenheim:2023izn}  specific form of path integral generated by a CQ proto action $W_{CQ}[\phi, q] =\int d^4 x \mathcal{W}_{CQ}[\phi, q]$ was introduced. For a single classical scalar field, the action takes the form
\begin{equation}\label{eq: CQProto}
\begin{split}
    &   \mathcal{I}(  \q, \xqr,  \xql ) = \int d^4 x  \big[i \mathcal{W}_{CQ}^+ -  i \mathcal{W}_{CQ}^- \\
    & -\frac{1}{2} \frac{\delta \Delta W_{CQ}}{\delta q} D_0(q) \frac{\delta \Delta W_{CQ}}{\delta q}  - 2 \frac{\delta \bar W_{CQ}}{\delta q} D_0(q)\frac{\delta \bar W_{CQ}}{\delta q} \big],
\end{split}
\end{equation}
where $D_0>0$, enforcing that the path integral converges. In Equation \eqref{eq: CQProto}, $W_{CQ}[\q,\psi]$ is a real classical-quantum proto-action which generates the dynamics, and we have introduced the notation $\bar{W}_{CQ} = \frac{1}{2 }( W_{CQ}[q,\psi^+] + W_{CQ}[q,\psi^-])$ for the $\pm$ bra-ket averaged proto-action and $\Delta W_{CQ} =  W_{CQ}[q,\psi^-] - W_{CQ}[q,\psi^+]$ for the difference in the proto-action along the $\pm$ branches. 
This form of action is motivated by the study of path integrals \cite{UCLPILONG} for CQ master Equations whose back-reaction is generated by a Hamiltonian \cite{Diosi:2011vu, Oppenheim:2018igd, UCLHealing}, as well as the purely classical path integral \cite{Weber_2017, Kleinert}. 

Written in the form of Equation \eqref{eq: CQProto}, we see that the action of the $\frac{\delta \bar W_{CQ}}{\delta q} D_0 \frac{\delta \bar W_{CQ}}{\delta q}$ term is to suppress paths that deviate from the $\pm$ averaged Euler-Lagrange equations, which themselves follow from varying the bra-ket averaged proto-action $\bar{W}_{CQ}$. The diffusion of the paths away from their averaged equations of motion is thus described by the diffusion coefficient by $D_0^{-1}$. The effect of the $\frac{\delta \Delta W_{CQ}}{\delta q} D_0 \frac{\delta \Delta W_{CQ}}{\delta q}$ term is to decohere the quantum system.

The decoherence diffusion trade-off \cite{dec_Vs_diff} means that maintaining coherence means there is necessarily diffusion in the classical system away from its most likely path, with the amount depending on both $D_0$ and the strength of the CQ coupling, which enters in $W_{CQ}$. The path integral in Equation \eqref{eq: CQProto} preserves purity on the quantum system when conditioned on the classical degree of freedom. This can be seen by expanding out the path integral and noting the lack of $\pm$ cross terms \cite{Oppenheim:2023izn}, meaning that once conditioned on the classical dynamics, the path integral factorizes into separate bra and ket integrals implying a pure quantum state evolution.  Explicitly, expanding out the cross terms in Equation \eqref{eq: CQProto} the action takes the form 
\begin{equation}\label{eq: CQProto2}
\begin{split}
    &   \mathcal{I}(  \q, \xqr,  \xql ) = \int d^4 x  \big[i \mathcal{W}_{CQ}^+ -  i \mathcal{W}_{CQ}^- \\
    & -\frac{D_0(q)}{2} \big(\frac{\delta  W_{CQ}^+}{\delta q}\big)^2 -\frac{D_0(q)}{2} \big(\frac{\delta  W_{CQ}^-}{\delta q}\big)^2   \big].
\end{split}
\end{equation}
The fact quantum states can remain pure is also shown in \cite{UCLHealing} using master-equation methods. We will often use Equation \eqref{eq: CQProto2} since it allows us to seperate the bra and ket branches. However, Equation \eqref{eq: CQProto} makes the consequences of the dynamics and the corresponding decoherence and diffusion evident.

One must further ensure that $\eqref{eq: CQProto}$ is normalized, and we elaborate on this point in Appendix \ref{section: normalizing}. When the CQ coupling does involve any a higher derivative kinetic terms, it was shown in \cite{UCLPILONG} that the dynamics generated by Equation \eqref{eq: CQProto} is normalized. In Appendix \ref{section: normalizing}, we extend this result and show that any CQ path integral of the form 
\begin{equation}\label{eq: normalizedCQmainLower}
\begin{split}
    & I[q,\psi^+, \psi^-]= \int dt i \dot{\psi}_+^2 + i V(\psi^+) - i \dot{\psi}_+^2 -i V(\psi^-) \\
    &- \frac{D_0(q,\dot{q}, \psi^+)}{2}(\ddot{q} + f(q,\dot{q}, \psi^+))^2 \\
    & - \frac{D_0(q,\dot{q}, \psi^-)}{2}(\ddot{q} + f(q,\dot{q}, \psi^-))^2,
    \end{split}
\end{equation}
is normalized. In the case $D_0>0$ has a functional dependence on the fields, one must make sure to also include a factor of  $\sqrt{\det(D_0(q,\dot{q}, \psi)}$ in the path integral measure \cite{UCLPILONG,Onsager1953Fluctuations,Dekker}. We further show that any higher derivative CQ path integral of the form 
\begin{equation}\label{eq: normalizedCQmain}
\begin{split}
    & I[q,\psi^+, \psi^-]= \int dt i \ddot{\psi}_+^2 + i V(\psi^+, \dot{\psi}^+) - i \ddot{\psi}_+^2 -i V(\psi^-, \dot{\psi}^-) \\
    &- \frac{D_0(q,\dot{q}, \psi^+, \dot{\psi^+})}{2}(\ddot{q} + f(q,\dot{q}, \psi^+, \dot{\psi}^+))^2 \\
    & - \frac{D_0(q,\dot{q}, \psi^-,\dot{\psi^-})}{2}(\ddot{q} + f(q,\dot{q}, \psi^-, \dot{\psi}^-))^2,
    \end{split}
\end{equation}
is also normalized up to constant factors. In the case $D_0>0$ has a functional dependence on the fields, one must again include a factor of  $\sqrt{\det(D_0(q,\dot{q}, \psi,\dot{\psi})}$ in the path integral measure \cite{UCLPILONG,Onsager1953Fluctuations,Dekker}. Note, if $D_0$ has a functional dependence on the fields, it is possible to re-exponentiate $\sqrt{\det(D_0)}$ through a Faddeev-Popov type action \cite{Bastianelli_2017}
\begin{equation}\label{eq: Fadeev}
S[a,\bar{b}, b]= \int dt D_0( a^2 + \bar{b} b),
\end{equation}
where $a$ is bosonic and $b,\bar{b}$ are anti-commuting Fermions. The integral over $b,\bar{b}$ yields $\det (D_0)$, whilst the integral over $a$ yields $(\det (D_0))^{-1/2}$ \cite{Bastianelli_2017}. We do not consider this explicitly here. 

Equation's \eqref{eq: normalizedCQmainLower} and \eqref{eq: normalizedCQmain} are very generic type of action one gets through Equation's \eqref{eq: CQProto} and \eqref{eq: CQProto2} when varying a CQ proto action that has second order equations of motion in the classical degree of freedom.

\subsection{Recovering measurement probabilities from CQ dynamics}\label{sec: Born}
In this section, we show how measurement probabilities are recovered by taking a  $D_0 \to \infty$ of the proto-action in Equation \eqref{eq: CQProto}. For the quantum gravity model we consider in Section \ref{sec: QGScalar}, this limit is similar to considering superpositions of spacetimes with large Ricci scalar. CQ collapse dynamics has previously been studied via unraveling approaches in \cite{UCLHealing, oppenheim2020objective}.

For simplicity, we consider a simple quantum mechanical model generated by the proto action 
\begin{equation}\label{eq: protoBorn}
    W_{CQ}[q,\psi] = \int dt \frac{\dot{q}^2}{2m_q} + \mathcal{L}(\psi) - \frac{k}{2}(q-\psi)^2.
\end{equation}
In Equation \eqref{eq: protoBorn}, $q$ is the classical particle, $\psi$ is the quantum particle, $k$ determines the strength of the CQ interaction, and $\mathcal{L}(\psi)$ is a purely quantum Lagrangian given by $\mathcal{L}(\psi) = \frac{\dot{\psi}^2}{2} + V(\psi)$. Equation \eqref{eq: protoBorn} will correspond to a $\psi$ measurement because the decoherence term takes the form $-\frac{D_0}{2}(\psi^+-\psi^-)^2$. 

With the proto action in Equation \eqref{eq: protoBorn}, the CQ action can be read off from Equation \eqref{eq: CQProto}
\begin{equation}\label{eq: BornAction}
\begin{split}
I(  \q, \xqr,  \xql ) & = \int dt \ i \mathcal{L}^+ - i \mathcal{L}^+ - \frac{D_0 k^2}{2}( \psi^+-\psi^-) \\
& - 2D_0 k^2 ( \frac{\ddot{q}}{m_q k} - (q-\frac{\psi^+}{2} - \frac{\psi^-}{2}))^2 .
\end{split}
\end{equation}
Because the classical action is higher derivative, we must specify both $q$ and $\dot{q}$ for the CQ state. In particular we should specify boundary conditions for both the $q$ and $\dot{q}$ variables in the path integral \cite{Hawking_2002}. We discuss this in detail when considering the normalization of path integrals in Appendix \ref{section: normalizing}. 

 The parameter $\tilde{D}_0 = D_0 k^2$ acts like the strength of a weak continuous measurement \cite{2006Jacob}. In a weak continuous measurement, the standard Born rule probabilities are recovered in the limit $\delta t \to 0$, $\tilde{D}_0 \delta t \to \infty$ \cite{GisinCollapse} and we take the same limit in this section. 

In Equation \eqref{eq: BornAction}, taking $\tilde{D}_0 \delta t \to \infty$ has multiple effects. Firstly, consider the kinetic terms $\frac{\dot{\psi}^2}{2}$ that arise in the purely quantum Lagrangian $\mathcal{L}(\psi)$. For $\delta t \to 0$, these are highly oscillating terms that will average the integrand to zero unless $\psi^+_f=\psi^+_i$ and $\psi^-_f=\psi^-_i$. Therefore, the kinetic terms force the initial and final bra and ket states to be equal. Now consider the decoherence and diffusion terms in Equation \eqref{eq: BornAction}. In the strong measurement limit, the decoherence term acts to impose a delta function $\delta(\psi^+_i- \psi^-_i)$ on the initial bra and ket. In combination with the kinetic term, this has the effect of decohering the quantum state into the $\psi$ basis. On the other hand, the diffusion term enforces a delta function over classical paths 
\begin{equation}
 \delta( \frac{\ddot{q}}{m_q k } - (q-\frac{\psi^+_i}{2} - \frac{\psi^-_i}{2})).
\end{equation}
In total, we find that taking the limit $\delta t \to 0$, $\tilde{D}_0 \delta t \to \infty$ maps any initial CQ state to
\begin{equation}\label{eq: finalstatemea}
\begin{split}
& \cqstate(\dot{q}, \ddot{q}, \psi, \psi, t+\delta t)= \\
& \times \frac{1}{2m_q k} \int dq  \delta( \frac{\ddot{q}}{m_q k } - (q-\psi))  \cqstate(q,\dot{q}, \psi, \psi,t),
\end{split}
\end{equation}
with all other components vanishing. 

In Equation \eqref{eq: finalstatemea}, we note that the CQ state at $t+ \delta t$ is specified in terms of $\dot{q}$ and $\ddot{q}$ of the classical path, where we integrate over $q$. It is useful to re-scale  $\frac{\ddot{q}}{m_q k } \to \ddot{q}'$ in Equation \eqref{eq: finalstatemea}. Taking into account the change in probability measure, we find Equation \eqref{eq: finalstatemea} can be written

\begin{equation}\label{eq: finalstatemeas}
\begin{split}
& \cqstate(\dot{q}, \ddot{q}', \psi, \psi, t+\delta t)=  \int dq  \delta( \ddot{q'}- (q-\psi))  \cqstate(q,\dot{q}, \psi, \psi,t).
\end{split}
\end{equation}

 We can also write Equation \eqref{eq: finalstatemeas} in terms of the Hilbert space density matrix 
\begin{equation}\label{eq: deltaCQ}
\begin{split}
\cqstate(\dot{q}, \ddot{q}', t+\delta t)=\int d\psi dq  \delta( \ddot{q}' - (q-\psi)) \cqstate(q, \dot{q}, \psi, \psi,t)|\psi\rangle \langle \psi|.
\end{split}
\end{equation}
Importantly, the dynamics of the classical particle uniquely determines $\psi$ and vice-versa. In particular, performing the delta function integration in Equation \eqref{eq: deltaCQ}, we can write $\psi_{\ddot{q}} = q-\ddot{q}'$ as the $\psi$ that one would find given $\ddot{q}$ at the second time step. This enables us to write the final state as 
\begin{equation}
\cqstate(\dot{q},\ddot{q}', t+\delta t) = \int dq \cqstate(q, \dot{q}, \q-\ddot{q}',q-\ddot{q}',t)|q-\ddot{q}'\rangle \langle q-\ddot{q}'|,
\end{equation}
which we can also write as 
\begin{equation}\label{eq: born0}
\cqstate(\dot{q},\ddot{q}', t+\delta t)=  \int dq \Tr[ |q-\ddot{q}' \rangle \langle q-\ddot{q}'| \cqstate(q,\dot{q},t) ] |q-\ddot{q}' \rangle \langle q-\ddot{q}'|.
\end{equation}
Equation \eqref{eq: born0} is the CQ state one would find from applying the Born rule for a measurement in the $\psi$ basis, where the outcome $|q-\ddot{q}' \rangle$ is found, and where we also average over the classical $q$ variable. In particular, the quantum state that is found is encoded by the dynamics of $q$, and the stochasticity arises because the dynamics of $q$ is stochastic. Equation \eqref{eq: born0} is normalized since
\begin{equation}
\begin{split}
    & \int d\dot{q} d\ddot{q}'\Tr[\cqstate(\dot{q},\ddot{q}',
    t+\delta t)]  =\\
    & \int d q d\dot{q} d\ddot{q}'\Tr[ |q-\ddot{q}' \rangle \langle q-\ddot{q}'|\cqstate(q,\dot{q},t) ] =1.
\end{split}
\end{equation}

In quantum theory, usually we are interested in applying the Born rule to a quantum system that is not initially correlated with a classical one. In this case, the initial state takes the form $\cqstate(q,\dot{q},t) = p(\dot{q}, q,t) \sigma(t)$, and the final CQ state is found to be
\begin{equation}
\begin{split}
 & \cqstate( \dot{q}, \ddot{q}\,t+\delta t)= \int dq \Tr[ |q-\ddot{q}' \rangle \langle q-\ddot{q}'| \sigma(t) ] \\
 & \times p(q,\dot{q},t)|q-\ddot{q}' \rangle \langle q-\ddot{q}'|.
 \end{split}
\end{equation}
Integrating over the classical degrees of freedom, we find the final quantum state $\sigma(t+\delta t) =\int d\dot{q} d\ddot{q}'\cqstate(\dot{q}, \ddot{q},t+\delta t)$ can be written as 
\begin{equation}
\begin{split}
   \sigma(t+\delta t)  = \int d\psi \Tr[ |\psi \rangle \langle \psi| \sigma(t) ] |\psi \rangle \langle \psi|,
\end{split}
\end{equation}
and so we find the Born rule of probabilities for a measurement of $\psi$. The probabilities of the measurement occur due to the stochasticity in the classical degree of freedom. In the $\tilde{D_0} \delta t \to \infty$ limit, for each stochastic classical trajectory we find a unique quantum state, with probability given exactly by the Born rule of probabilities. In this way CQ theories are able to avoid contradictory behaviour between unitary and non-unitary evolution; there is only dynamics of interacting classical and quantum systems, and the dynamics is non-unitary always. 

Note, that the imposition of the $\tilde{D}_0 \delta t \to 0$ means that the classical system can also act to measure entangled states. This is an important and subtle point, since CQ dynamics is linear in the state, and has also been shown to be equivalent to measurement and feedback \cite{UCLHealing}. Let us explore this further by considering a CQ state of a joint system $A,B$. We apply the CQ measurement described by Equation \eqref{eq: BornAction} on system $A$ alone. In this case, the state is described by
$\cqstate(q_A,\dot{q}_A,\psi^{\pm}_A, \psi^{\pm}_B)$. 
The dynamics defined by Equation \eqref{eq: BornAction} is also local, and acts to impose the delta functions $\delta(\psi^+_A-\psi^-_A)$ and $\delta( \frac{\ddot{q}_A}{m_q k } - (q_A-\frac{\psi^+_A}{2} - \frac{\psi^-_A}{2}))$ on the $A$ system alone. However, the CQ entangled  $\cqstate(q_A,\psi^{\pm}_A, \psi^{\pm}_B)$ state encodes non-local correlations. The local delta functions then act on the state according to the Born rule on $A$ alone, and change the global structure of the state. In analogy with classical probability theory, the imposition of the delta functions then acts as a kind of dynamic conditioning on the combined CQ state. This conditioning leads to global changes in the entangled state because it is non-separable and encodes Bell non-local correlations \cite{Bell:1964kc}.

\section{Quantum gravity with dynamical collapse}\label{sec: QGScalar} We now consider a classical scalar field coupled to quantum gravity. We denote the classical scalar field by $\phi$, and the quantum degrees of freedom will be the spacetime metric $g_{\mu \nu}$. We let these interact dynamically so that there is both an effect of the classical scalar field on the gravitational metric and a back-reaction of the quantum spacetime on the classical scalar field. To have such a back-reaction and maintain complete positivity, the dynamics are non-unitary but instead described by the CQ dynamics introduced in the previous section. 

We take a path integral description, which takes the form of a sum over the paths of a classical field and the quantum metric. The partition function then takes the form 
\begin{equation}\label{eq: partition}
\begin{split}
    Z &=\int^{B} \mathcal{D}\phi \mathcal{D}g_{\mu\nu}^+ \mathcal{D}g_{\mu \nu}^-  \mathcal{D}\psi^+ \mathcal{D}\psi^-  \exp( I[\phi, g_{\mu \nu}^+ , g_{\mu \nu}^-, \psi^+, \psi^-]) \\
    & \times \cqstate(\phi, g_{\mu \nu}^+, g_{\mu \nu}^-, -\infty)
\end{split}
\end{equation}
where the boundary conditions impose that
\begin{equation}
    B =\{ \psi^+(\infty) =\psi^-(\infty), g_{\mu \nu}^+(\infty) = g_{\mu \nu}^+(\infty)\},
\end{equation}
and $\psi$ is a schematic representation of the remaining quantum matter fields. We take the action $I$ in Equation \eqref{eq: partition} to be of the form in Equation \eqref{eq: CQProto} so that the dynamics are completely positive. We do not attempt to rigorously justify how to define the path integral over spacetime metrics, nor a detailed consideration of gauge invariance, but study its weak field limit in Section \ref{sec: peturbative} using standard methods \cite{Donoghue:2012zc, Veltman:1975vx}. 

In Equation \eqref{eq: partition}, we also make no attempt to study precisely the asymptotic steady states of the CQ system, and our main discussion pertains to the CQ action. For now, we simply mention that since we consider a CQ coupling only to gravity and not matter, one can imagine that the decohered Minkowski metric is stable, along with the energy eigenstates of matter. This is plausible since matter weakly interacts with gravity, and we do not consider any changes to the matter sector. Clarifying the steady states of CQ systems and how one computes scattering amplitudes in practice, we leave to future work.

\subsection{Coupling gravity to a Brans-Dicke scalar field}
Let us now motivate the form of coupling that gives rise to an appropriate amplification mechanism,  whereby unitarity is approximate for inherently quantum states but not for macroscopic ones. For this to occur, we would like the classically induced decoherence term, given by $ \frac{\delta \Delta W_{CQ}}{\delta \phi}$, to have a dependence (directly or indirectly) on the stress-energy tensor of matter or the Ricci curvature. 

As such, a natural starting point is then to search for a proto-action $W_{CQ}$ where the classical equations of motion for a scalar field $\phi$, found by varying $\frac{\delta W_{CQ}}{\delta \phi}$, lead to equations of the form $\Box \phi = \lambda T$, where $T$ is the trace of the stress-energy tensor. In this way, the decoherence arising from Equation \eqref{eq: CQProto} will be related to the stress-energy tensor of matter.

Classically, such equations of motion arise from non-minimal Brans-Dicke couplings \cite{Brans_1961, Kofinas:2015sjz} of the form
 \begin{equation}\label{eq: Brans-Dicke}
 \begin{split}
    S_{BD}[\phi,g] & =  \int d^4 x \sqrt{-g}\frac{1}{16 \pi }( \phi R - \frac{\omega}{\phi} \nabla_a \phi \nabla^a \phi)  + \mathcal{L}_m,
     \end{split}
 \end{equation}
 where $\omega$ is a coupling constant, $R$ is the Ricci-scalar, which we will later take to be associated with a quantum spacetime, and $\mathcal{L}_{m}$ denotes the matter Lagrangian. 
 
 The Brans-Dicke coupling can be understood as promoting Newton's constant $\frac{1}{G}$ to a dynamical scalar field $\phi$. Extensions of the Brans-Dicke models with a functional dependence $\omega(\phi)$ have also been considered \cite{Bergmann1968, Wagoner1970,Roy2017}. For simplicity, we focus on the case $\omega$ is constant. We also mention that it is possible to consider a plethora of other non-minimal couplings between a classical scalar field and gravity, such as the Horndeski theories \cite{Horndeski1974}.
 
For the purposes of this work, we find the Brans-Dicke theories the most computationally simple. As a competitor to general relativity, the Brans-Dicke theory is consistent with observations for $\omega > 40,000$ \cite{Will2014, Bertotti2003}, but the largeness of the parameter is often thought unnatural; this problem is also present here and is perhaps why more involved couplings such as those contained in the Horndeski theories may be preferred. 
 
 The classical equations of motion for the action in Equation \eqref{eq: Brans-Dicke} are found by variation of the action with respect to the scalar field 
 \begin{equation}\label{eq: SFBD}
 \begin{split}
    \frac{16 \pi}{\sqrt{g}} \frac{S_{BD}[\phi,g]}{\delta \phi} = -R - \frac{2 \omega}{\phi} \Box \phi + \frac{\omega}{\phi^2} \nabla_a\phi \nabla^a \phi,
     \end{split}
 \end{equation}
 and the metric
  \begin{equation}\label{eq: EEBD}
 \begin{split}
    \frac{16 \pi}{\sqrt{g}}  \frac{S_{BD}[\phi,g]}{\delta g^{\mu \nu}} & = 8 \pi T_{\mu\nu} + \frac{\omega}{\phi} \left( \phi_{,\mu} \phi_{,\nu} - \frac{1}{2} g_{\mu\nu} \phi_{,\alpha} \phi^{,\alpha} \right) \\
    & + \phi_{,\mu; \nu} - g_{\mu \nu} \Box \phi - \phi G_{\mu \nu}.
     \end{split}
 \end{equation}
 On shell, Equation's \eqref{eq: SFBD} and \eqref{eq: EEBD} vanish. Taking the trace of \eqref{eq: EEBD} and substituting it into Equation \eqref{eq: SFBD}, we find the equation of motion of the scalar field
 \begin{equation}\label{eq: phiEM}
     \Box \phi = \frac{8 \pi T }{( 3+ 2\omega)}.
 \end{equation}
In the limit $\omega \to \infty$, the solution to Equation \eqref{eq: phiEM} is $\phi = \phi_0$, and taking $\phi_0 = \frac{1}{G}$, Equation \eqref{eq: EEBD} leads to the standard Einstein's equations. 

 Let us now consider the CQ theory generated by the Brans-Dicke action. Because of Equation \eqref{eq: normalizedCQmain}, we must also add quadratic kinetic gravity terms in the CQ proto action to ensure the CQ theory is normalized. We further expect that terms in action that do not appear with derivatives, i.e, the lapse and shift vectors, are treated as gauge parameters that need to be fixed by a Faddeev-Poppov method. These should also multiply constraints, and we leave a study of the constraints of the theory to future work. With the addition of quadratic gravity terms, we take the CQ proto action to be
 \begin{equation}\label{eq: BDNorm}
 W_{CQ} = S_{BD} + S_{HD}
\end{equation}
where 
\begin{equation}
  S_{HD} = \int d^4 x \mathcal{L}_{HD}=  \int d^4x \sqrt{-g} \left( - \beta R^2 + \alpha R_{\mu\nu} R^{\mu\nu} \right),
\end{equation}
are the Higher derivative quadratic gravity terms. 

For ease of calculation, and so we can give a gentler introduction to the theory, we will not consider the effect of the quadratic gravity terms explicitly -- except when considering details of renormalization in Section \ref{sec: renormalization}. Indeed, we do not couple these terms to the classical field, so much of our presentation can be understood as the limit where $\alpha, \beta$ are small. Nonetheless, including the quadratic Ricci terms in Equation \eqref{eq: BDNorm} is important and another way the theory will deviate from general relativity in the low energy limit. When viewed as an effective theory, more generally we expect the action to include a tower of terms higher order in the spacetime curvature \cite{Donoghue_1994} (see Section \ref{sec: effective}).

Substituting for the action in Equation \eqref{eq: CQProto2}, and absorbing factors of $(16\pi)^{-2}$ into $D_0$, we find the CQ action $I[\phi, g_{\mu \nu}^+, g_{\mu \nu}^-] = I_{CQ}[\phi, g_{\mu \nu}^+] +I_{CQ}^*[\phi, g_{\mu \nu}^-] $,
where
 \begin{equation}\label{eq: CQBD}
 \begin{split}
    &  I_{CQ}[\phi, g_{\mu \nu}] = \\
    & \int d^4 x  \big[i\sqrt{-g}(\frac{\phi R}{16 \pi} + \mathcal{L}_{HD} +\mathcal{L}_m - \frac{\omega}{\phi} \partial_a\phi \partial^a \phi )\\
     & - \frac{1}{2}D_0 \sqrt{-g}( \frac{2 \omega}{\phi} \Box \phi - \frac{\omega}{\phi^2} \partial_a\phi \partial^a \phi + R) ^2.
 \end{split}
 \end{equation}
We remind the reader that the $\pm$ superscript refers to the evaluations on the bra and ket branches, and that we have the CQ action in Equation \eqref{eq: CQBD} for both the bra and ket branches. For simplicity, we take $D_0=const$. Equation \eqref{eq: CQBD} describes a quantum-gravity density matrix interacting with a classical field. The $\sqrt{-g^{\pm}}$ factors appearing in the bra-ket path integral mean that quantum states in superposition of spacetimes will pick up different phases due to their classical evolution since it couples to $\sqrt{-g}$. 
 
 We chose the Brans-Dicke coupling for the proto action because it was the simplest form of coupling that gives rise to a collapse mechanism whose classical limit is generated by an action. However, there is no need for the CQ action to be generated by a proto-action. We could, therefore, start from Equation \eqref{eq: normalizedCQmain} directly, which defines normalized completely positive CQ dynamics. In doing so, we are able to write down much simpler sets of theories that maybe more suitable for gaining a further understanding of collapse dynamics. As an example, we can consider the dynamics of a non-minimally coupled scalar field $\phi$
 \begin{equation}\label{eq: CQBDSimple}
 \begin{split}
    &  I_{CQ}[\phi, g_{\mu \nu}] = \\
    & \int d^4 x  \big[i\sqrt{-g}(\frac{R}{16 \pi G}(1- \xi \phi^2) + \mathcal{L}_{HD} +\mathcal{L}_m)\\
   & \frac{1}{2}D_0 \sqrt{-g} (\Box \phi  - \lambda R)^2 ,
 \end{split}
 \end{equation}
 where $\xi$ is the coupling. Equation \eqref{eq: CQBDSimple} has a very similar weak field to the Brans-Dicke theory but is much simpler. Its classical limit is not generated by an action, even though the full dynamics is described by a combined classical-quantum action. Nonetheless, simpler forms of the dynamics such as Equation \eqref{eq: CQBDSimple} might be a useful starting place to study toy models. 
 
 To gain insight into Equation \eqref{eq: CQBD} we now study the weak field limit, where we perturb around a decohered background which we take to be Minkowski. 

\subsection{Weak field limit}\label{sec: peturbative}
In this section, we study the weak field limit of the CQ theory so that we can gain insight into the low energy predictions of the theory. 

In the weak field approximation of the CQ Brans-Dicke theory we decompose the bra-ket metric fields and the scalar fields according to 

\begin{equation}\label{eq: expansion}
\begin{split}
    & g_{\mu \nu}^+ = \eta_{\mu \nu} + h_{\mu \nu}^+, \\
      & g_{\mu \nu}^- = \eta_{\mu \nu} + h_{\mu \nu}^-, \\
    & \phi = \phi_0 + \epsilon,
\end{split}
\end{equation}
which can be understood as expanding around a decohered Minkowski background.

Inserting the expansion of Equation \eqref{eq: expansion} into the action of Equation \eqref{eq: CQBD}, and keeping only quadratic terms, we find the full CQ action takes the form 
\begin{equation}\label{eq: fullWF}
 \begin{split}
      I[\epsilon, h_{\mu \nu}^+, h_{\mu \nu}^-] &= \int d^4 x \sqrt{-\eta} \big[i S[\epsilon, h_{\mu\nu}^+]-iS[\epsilon, h_{\mu\nu}^-] \\
     & - \frac{1}{2}D_0(R^+_{(1)}-R^-_{(1)})^2 \\
     & - \frac{8D_0 \omega^2}{\phi_0^2}( \Box \epsilon + \frac{\phi_0}{4 \omega}(R^+_{(1)} + R^-_{(1)})) ^2 \big],
 \end{split}
 \end{equation}
 where 
 \begin{equation}
    R_{(1)} = \partial_{\mu} \partial_{\nu} h^{\mu \nu} - \Box h, 
\end{equation}
is the linearized Ricci-Scalar. In Equation \eqref{eq: fullWF} $S_0[\epsilon, h_{\mu\nu}]$ is the unitary part of the evolution - which describes linearized Brans-Dicke quantum gravity - and is defined by
\begin{equation}\label{eq: unitaryBD}
   S_0[\epsilon, h] = i(\phi_0 \mathcal{L}_{LG} + \epsilon R_{(1)} + \mathcal{L}_{HD} +  \mathcal{L}_m + h_{\mu\nu}T^{\mu\nu }),
\end{equation}
where $\mathcal{L}_{LG}$ is the linearized gravity Lagrangian, $\mathcal{L}_{HD}[h]$ is the linearized quadratic gravity action \cite{Salvio_2018}, $T_{\mu\nu}$ is the stress energy tensor of matter, and $\mathcal{L}_m$ is now evaluated in the Minkowski background.

Let us now go through each of the terms appearing in Equation \eqref{eq: fullWF}. The first line of Equation \eqref{eq: fullWF} describes the purely unitary quantum  gravity dynamics of the system. Through Equation \eqref{eq: unitaryBD}, we see that Newton's constant is replaced by the classical Brans-Dicke field. There is also a coupling of the Ricci scalar to $\epsilon$. Because of the appearance of the classical field, this term can be understood as unitary evolution with classical control.
 
 The second line of Equation \eqref{eq: fullWF} acts to decohere quantum states according to their Ricci scalar. Intuitively, macroscopic states have larger Ricci scalar and so decohere faster. If $D_0$ is chosen appropriately, this term reflects that the dynamics are approximately unitary on small scales but inherently classical for macroscopic objects. As such, a measurement postulate isn't necessarily required, since the classical field acts to measure spacetimes according to their Ricci scalar. We study the decoherence rate in more detail when we study the low energy predictions of the theory in Section \ref{sec: peturbative}. 
 
 The final line of Equation \eqref{eq: CQBD} describes the back-reaction of the quantum gravitational field on the classical scalar field, with associated diffusion parameterized by the inverse of $D_0$. In other words, the  CQ coupling means that the classical scalar field diffuses around the $\pm$ average of its classical solution, closely related to the quantum expectation value of its equations of motion \cite{UCLHealing}. In the weak field limit, this diffusion manifests as diffusion in the Newtonian coupling, as we will show via Equation \eqref{eq: diffusionNP}. The observation of decoherence and diffusion is a general prediction of CQ theories \cite{dec_Vs_diff, Galley:2023byb}.
 
 Expanding out Equation \eqref{eq: fullWF}, we can also write it in terms of 
 \begin{equation}
     I = I_{CQ}^+ + I_{CQ}^{*-} - I_C,
 \end{equation}
 where

\begin{equation}\label{eq: CQonlyWeak}
 \begin{split}
      I_{CQ}[\epsilon, h]  & =\int d^4 x \sqrt{-\eta} \big[i(\phi_0 \mathcal{L}_{LG}+ \epsilon R_{(1)} + \mathcal{L}_m \\
     & - D_0 R_{(1)}^2 - 4D_0 \omega R_{(1)}\frac{\Box \epsilon}{\phi_0} \big],
 \end{split}
 \end{equation}
 and
 \begin{equation}\label{eq: WFClass}
 I_{C}[\epsilon]  =\int d^4 x \sqrt{-\eta} \frac{8D_0 \omega^2}{\phi_0^2}( \Box \epsilon)^2.
 \end{equation}
 
In Equation \eqref{eq: fullWF}, the combined CQ dynamics is written in the language of a coupled quantum field theory, where there is also a classical field that acts as a classical probability measure. It is not of standard form since it is not unitary, though we emphasize that the CQ path integral is designed to be completely positive, so the non-unitarity of the quantum sector can be consistent with positive probabilities. In particular, the bra and ket path integrals are indirectly coupled via the classical scalar field.

Though the theory we discuss is non-unitary, we note that bra-ket couplings have recently been studied to resolve paradoxes that arise in gravitational path integrals \cite{Chen:2020tes, Almheiri_2020, penington2020replica}. In particular, the replica geometries introduced to reproduce the Page curve of an evaporating black hole \cite{ Almheiri_2020, penington2020replica}  couple bra and ket variables. Because of this, it has been argued that they should be interpreted as computing an ensemble average \cite{penington2020replica, stanford2020quantum}. It would be interesting to explore the parallels here. For the CQ theory, the fine grained entropy's computed conditioned on the classical variable are not the same as the coarse grained entropy's found by computing the entropy of the classically averaged state because the replicas are coupled. This is a well known phenomena often studied in measurement-induced phase transitions \cite{Buchold_2021}.

 \section{Low energy predictions of the theory}\label{sec: pred} 

In this section we discuss some general low-energy, non-relativistic predictions of the theory defined by Equation \eqref{eq: CQBD}. As we discuss in Section \ref{sec: renormalization}, we expect the theory should be viewed as an effective theory since it is not renormalizable. Hence, though we cannot trust its high energy predictions for Planck scale physics, we do expect to be able to trust low energy ones.

Extracting precise experimental predictions for the theory requires an in-depth study of the weak field limit through the lens of effective field theory. Therefore, we present heuristic derivations of macroscopic predictions that one might observe. Even if other theories are physically preferred, we expect that these are fairly generic predictions for any quantum gravity theory coupled to a scalar field: we expect that such a theory will end up predicting deviation from Einstein's equations through a non-minimal coupling, decoherence of the quantum spacetimes by an amount depending on its curvature, and diffusion in the scalar field, manifesting itself in terms of random sources to the Newtonian potential in the classical limit. Indeed, the behavior of decoherence and diffusion in classical-quantum theories is generic \cite{dec_Vs_diff, Galley:2023byb}, and has been used to put constraints on theories where gravity is classical \cite{dec_Vs_diff}. The precise details will, of course, depend on the coupling.  It would also interesting to explore the cosmological effects of such theories but we do not do this here. 

Let us know consider the consequences of Equation \eqref{eq: fullWF} in the low energy limit.
 
\noindent\textit{1. Deviations from GR through $\omega$ and quadratic gravity terms}. The theory deviates from Einstein's equations for finite values of $\omega$ \cite{Will2014} through the Brans-Dicke mechanism. The best current bound on Brans-Dicke couplings comes from Solar observations, placing $\omega>40,000$ \cite{Will2014, Bertotti2003}. The theory also predicts deviations from General relativity due to the higher derivative terms in Equation \eqref{eq: BDNorm}, but since these are not coupled to a classical field, we do not discuss them here. 

It is interesting to study the $\omega \to \infty$ in the CQ theory. In the $\omega \to \infty$ limit, Equation \eqref{eq: fullWF} becomes approximately
\begin{equation}\label{eq: approxinf}
    - \frac{1}{2 D_2}( \Box \epsilon + \frac{\phi_0}{4 \omega}(R^+_{(1)} + R^-_{(1)})) ^2 \approx - 4 D_0\omega^2( \Box \epsilon) ^2.
\end{equation}
Equation \eqref{eq: approxinf} then enforces that $\epsilon$ is constant. Hence, we can reabsorb it into $\phi_0$, and we find that the remaining path integral described standard quantum gravity with Newtonian constant $G = \frac{1}{\phi_0}$ but with the addition of a decoherence term $D_0( R^+ - R^-)^2 $. In this limit, the purity of the quantum system is not maintained unless the simultaneous limit $D_0 \to 0$ is also taken, in which case we recover standard perturbative quantum gravity.

\noindent\textit{2. Decoherence of quantum superpositions through $D_0$.} The theory predicts the decoherence of quantum superpositions at a rate defined by the difference in Ricci scalar between the two elements in supposition. For a quantum superposition of non-relativistic particles, the Ricci scalar is given by $R= - \nabla^2 \Phi =- 4 \pi G m(x)$, where $m(x)$ is the mass density of the particle. 

As a consequence, we expect that the decoherence rate for a mass of positions in superposition of $m_L,m_R$ states reads 
\begin{equation}\label{eq: decRate}
    \lambda \sim (4\pi G)^2 \int d^3x D_0 (m_L(x)-m_R(x))^2,
\end{equation}
which is the same formula for the decoherence rate posited in many of the non-relativistic collapse models \cite{Pearle:1988uh,PhysRevA.42.78,BassiCollapse,PhysRevD.34.470, GisinCollapse, relcollapsePearle, 2006RelCollTum}. However, instead of smearing $D_0$ over space, we expect that one controls the divergent contributions to the decoherence rate for point particles by considering the path integral as an effective action and treating the bra-ket correlation functions as a peturbative expansion dictated by the energy scales in the theory. In this way, the divergent contributions to the decoherence rate are tamed through cut-offs in Fourier space.

\noindent\textit{3. Diffusion of the effective Newtonian constant through $(D_0 \omega^2)^{-1}$.} The theory also predicts diffusion of the scalar field, which acts as a dynamical Newtons constant. To better understand the effect of diffusion, we now study the non-relativistic weak field limit of classical Brans-Dicke theory. The diffusion effects are then discussed by adding noise to the classical solution, described by Equation \eqref{eq: randomSF}. 

The weak field limit of classical Brans-Dicke is most easily studied in the \textit{Einstein frame} \cite{Barros1998}. To that end, we define 

\begin{equation}
    \bar{g}_{\mu\nu} = \frac{\phi}{\phi_0} g_{\mu\nu}, 
\end{equation}
where the bar in \( \bar{G}_{\mu\nu} \) and means that these quantities are now calculated using the unphysical metric \( \bar{g}_{\mu\nu} \).

Defining \( G_0 = \frac{1}{\phi_0} \), it is shown in \cite{Barros1998} that the weak field equations for a matter distribution $T_{\mu \nu}$ are given by
\begin{equation}\label{eq: WFmetric}
    \bar{G}_{\mu\nu} = 8\pi G_0 T_{\mu\nu},
\end{equation}
and 
\begin{equation}\label{eq: WFepsilon}
   \Box \epsilon = \frac{8 \pi T }{2\omega + 3}.
\end{equation}

The equation for the unphysical metric $\bar{g}_{\mu \nu}$ is hence formally identical to the field equations of General Relativity but with \( G_0 \) replacing the Newtonian gravitational constant \( G \). 

Thus, if \( \bar{g}_{\mu\nu}(G_0, x) \) is a solution of the Einstein equations in the weak field approximation for a given \( T_{\mu\nu} \), then the Brans-Dicke solution corresponding to the same \( T_{\mu\nu} \) is given by 
\begin{equation}\label{eq: unphymet}
    g_{\mu\nu}(x) = G_0^{-1} \phi^{-1} \bar{g}_{\mu\nu}(G_0, x) = [1 - \epsilon(x)G_0]\bar{g}_{\mu\nu}(G_0, x).
\end{equation}
As a consequence, in the weak field limit, solutions of Brans-Dicke equations of gravity are reduced to solving Einstein field equations for the same matter distribution and for the scalar field $\epsilon$ via Equation \eqref{eq: WFepsilon}.

For a point particle $T^{\mu}_{\nu} = m \delta(r) \delta^{\mu}_0 \delta^0_{\nu}$, where $m$ is the mass of the particle, and the standard Newtonian solution with Newtons constant $G_0$ is given by 
\begin{equation}
    ds^2 = \left(1 - \frac{2mG_0}{r}\right) dt^2 - \left(1 + \frac{2mG_0}{r}\right) \left(dr^2 + r^2d\Omega_2^2\right).
\end{equation}
The equation for the scalar perturbation $\epsilon$ becomes a Poisson equation
\begin{equation}
    \nabla^2 \epsilon = - \frac{8 \pi m}{2 \omega + 3 }\delta(r) ,
\end{equation}
which is solved via
\begin{equation}
    \epsilon = \frac{2m }{(2 \omega + 3) r}.
\end{equation}
Inserting the solution into Equation \eqref{eq: unphymet} one can read off $h_{00}$ as 
\begin{equation}
    h_{00} = -\frac{2mG_0}{r} \frac{2 \omega + 4}{2 \omega + 3} = - \frac{2m G}{r} = 2\Phi,
\end{equation}
where we have related the Brans-Dicke couplings $G_0$ and $\omega$ to $G$ by setting $\frac{1}{\phi_0 } = G_0 = \left( \frac{2 \omega + 3}{2 \omega + 4} \right) G $. 

The result of the CQ coupling is to diffuse around the classical solution for $\epsilon$. Assuming the matter is decohered, the diffusion is embodied by the Langevin equation \cite{dec_Vs_diff}
\begin{equation}\label{eq: randomSF}
    \Box \epsilon = \frac{8 \pi T  }{2 \omega + 3} + \big(\frac{\phi_0^2}{8 D_0 \omega^2}\big)^{1/2} J ,
\end{equation}
where $J$ is a white noise process satisfying $\langle J \rangle =0$ and 
\begin{equation}
    \langle J(x) J(y) \rangle = \delta(x,y).
\end{equation}
In Equation \eqref{eq: randomSF}, the coefficient $\sigma = \big(\frac{\phi_0^2}{8 D_0 \omega^2}\big)^{1/2}$ is chosen so that $\langle \sigma J(x) \sigma J(y) \rangle = 2 D_2 \delta(x,y)$ yields the correct diffusion coefficient appearing in the path integral in Equation \eqref{eq: WFClass}. Again, we expect that the singularity of the white noise process corresponds to the fact that the path integral must be renormalized as an EFT.

In the Newtonian limit $c\to \infty$, the solution to Equation \eqref{eq: randomSF} for a particle located at the origin is given by 
\begin{equation}
    \epsilon = \frac{2m}{(2\omega + 3) r} + \big(\frac{\phi_0^2}{8 D_0 \omega^2}\big)^{1/2}\int d^3 x \frac{J(x')}{4 \pi |x-x'|}.
\end{equation}
Hence, to leading order, we can read off the effective Newtonian potential as 
\begin{equation}\label{eq: diffusionNP}
      h_{00} =2\Phi_{eff} = - \frac{2m G }{r}  -  \big(\frac{1}{8 D_0 \omega^2}\big)^{1/2}\int d^3 x \frac{J(x')}{ 4\pi |x-x'|}.
\end{equation}
Consequently, the diffusion around the classical solution leads to a stochastic sourcing of the Newtonian potential. The theory predicts that the effective Newtonian constant diffuses around an average value over time due to the diffusion induced in $\epsilon$. We take this as an indication that the stochastic fluctuations in the classical field can be interpreted as classical versions of quantum vacuum fluctuations. Note that the stochastic fluctuations in Equation \eqref{eq: diffusionNP} look non-local. This arises from taking the $c\to \infty$ limit of the wave equation for $\epsilon$. 

In summary, the theory is described by two parameters, $D_0$ and $\omega$, which have three concrete and distinct effects. $D_0$ is related to the decoherence rate of masses in superposition through Equation \eqref{eq: decRate}. On the other hand, $\omega$ describes the deviation from the Einstein equations, while the combination  
\begin{equation}\label{eq: dec_vs_diff}
    D_2 = (8 D_0\omega^2)^{-1}
\end{equation}
quantifies the amount of diffusion in the value of the effective Newtonian constant $\epsilon$, which we expect leads to diffusion in the Newtonian potential through Equation \eqref{eq: diffusionNP}. 
In contrast to the CQ theories of gravity studied in \cite{Oppenheim:2018igd, Oppenheim:2023izn, dec_Vs_diff} where the theory is heavily constrained due to a determined coupling between classical and quantum gravity, in Equation \eqref{eq: dec_vs_diff} the strength of the coupling mediating the CQ interaction is a free parameter $\omega$ to be determined.

One drawback of the present theory is that there is seemingly no experimental restriction on how large $\omega$ can be. For fixed $D_0$, as $\omega$ increases, there is less observable diffusion in the Newtonian potential, and the quantum gravity sector approaches that of Einstein. This feature is both a blessing and a curse: having a large value of $\omega$ seems unnatural and there is no limit on how large $\omega$ can be. On the other hand, a large value of $\omega$ means the back-reaction of the quantum metric on the classical scalar field is weak, in which case, via Equation \eqref{eq: dec_vs_diff}, the observed diffusion will also be very small, even for small decoherence rates. We expect that by considering alternative forms of CQ couplings based on Horndeski's theories, one can arrive at different CQ theories with similar properties that may not suffer from such problems of naturalness.

We leave experimental considerations of such a theory to future work. This requires a full understanding of the effective theory described by Equation \eqref{eq: fullWF}, and a better understanding of alternate couplings to gravity. However, we note that the trade-off between decoherence and diffusion, given in terms of the strength of the back-reaction, has been used to place constraints on the classicality of gravity and is generic for CQ theories \cite{dec_Vs_diff}. We also mention that, just as for the predictions of classical-quantum gravity, which also predict stochastic fluctuations in the Newtonian potential \cite{dec_Vs_diff}, it is curious that different experiments to measure Newton's constant $G$ yield results with large uncertainty \cite{quinn2000measuring,gillies2014attracting,rothleitner2017invited}.

\section{renormalization and effective CQ field theories}\label{sec: renormalization}
In this section, we consider the problem of renormalization of the CQ theory in more detail. We end up concluding that the theory should be viewed as an effective field theory, describing the low energy limit of a UV complete theory; the UV complete theory could contain a fundamentally classical scalar field, or the classicality of the scalar field may also be a low energy effect. 

Since we include higher order Ricci scalar terms in the action - both through Equation \eqref{eq: BDNorm} and the CQ coupling - let us first discuss features of quantum gravity when higher order curvature terms are considered in the action.

\subsection{Quadratic gravity}
Let us first discuss the features of higher derivative quadratic gravity. For an overview, we recommend \cite{Donoghue_2019,Salvio2018}. The action for quadratic gravity takes the form 
\begin{equation}\label{eq: higherderiv}
  S_{QG} = \int d^4x \sqrt{-g} \left(\frac{R}{16 \pi G} - \beta R^2 + \alpha R_{\mu\nu} R^{\mu\nu} \right),
\end{equation}
where $\alpha, \beta$ are the higher order couplings to the Ricci tensor and Ricci scalar. Quadratic gravity has been shown to be renormalizable in full, and a discussion of the renormalizability of higher derivative gravity, including gauge invariance considerations, is given in \cite{Stelle_1977}. 

At tree level, the problems with quadratic gravity are most easily seen by considering the form of the propagator, which takes the schematic form \cite{Donoghue_2019, Stelle_1977}
\begin{equation}\label{eq: quadgravmain}
\begin{split}
  D(p^2)& = \frac{1}{p^2- \frac{p^4}{M^2}} = \frac{1}{p^2} - \frac{1}{p^2 -M^2},
  \end{split}
\end{equation}
where $M^2$ can be related to $\alpha$ appearing in Equation \eqref{eq: quadgravmain} through $M^2  \sim \frac{1}{\alpha \phi_0}$ but it does not depend on the $\beta$ coupling \cite{Donoghue_2019}. Strictly speaking $M^2$ is associated with a spin-2 mode, and the exact propagator also includes a scalar mode $m^2$, related to both the quadratic gravity couplings $\alpha, \beta$ through $m^2 = \frac{1}{\phi_0^2(3\beta -\alpha)}$ \cite{Stelle_1977}. 

 The problem with Equation \eqref{eq: quadgravmain} is that the relative minus sign in the propagator with the $M^2$ pole signals a ghost particle whose propagation is time-reversed from usual particles, and so there is creation of correlations at space-like separated regions. Alternatively, one can view the negative propagator as arising from a negatively normed state \cite{Stelle_1977}. The particle appears stable at the tree level, and so there are potential violations of unitarity, stability, and causality. It has been argued that these effects are reduced when one considers corrections to the self-energy, which picks up an imaginary component \cite{Donoghue_2019_stab}. However, because of the ghost, higher derivative gravity is often considered as an effective field theory in the region $p^2\ll M^2$, but breaks down at energies similar to the ghost mass \cite{Steinwachs_2020}. 

 \subsection{f(R) gravity}
When viewed as a fundamental theory, quadratic gravity is may not be stable due to the $\alpha$ coupling to the Ricci tensor in Equation \eqref{eq: higherderiv} \cite{Steinwachs_2020}. A simple and phenomenologically important extension of GR is to consider \( f(R) \) gravity \cite{Steinwachs_2020}. In $f(R)$ gravity, one includes considers the Lagrangian to be an arbitrary function \( f \) of the Ricci scalar \( R \). For our purposes, we can view this as Equation \eqref{eq: higherderiv} with $\alpha =0$. This was first model of inflation and is in favour with Planck data \cite{2020_planck, Steinwachs_2020}.

Unfortunately, $f(R)$ theories of gravity are non-renormalizable. In particular, the one loop divergences have been calculated for perturbations around an arbitrary background \cite{Ruf:2017bqx}. The divergence structure requires counter terms of the form $R_{\mu \nu} R^{\mu \nu}$ that are not removable by counter terms of the form $f(R)$. As a consequence, $f(R)$ gravity must also be viewed as an effective theory valid only in the low energy regime. Viewed as an effective field theory, one starts with the standard gravitational propagator, considering the higher order terms in a perturbation series that are only valid below a certain energy scale. 

\subsection{The CQ gravity theory}
In the classical-quantum theory defined by Equation \eqref{eq: BDNorm}, the evolution on the quantum sector alone is non-unitary, and probabilities are conserved when the interaction with the classical sector is considered \cite{UCLHealing}. In the present case, the non-unitarity of the quantum sector is represented by an imaginary higher momentum contribution to the propagator. This can be viewed as a complex contribution to the $\beta$ coupling to $R^2$. In this convention, we include an overall $i$ in the quantum action, and the CQ couplings are real. The CQ couplings then add complex components to the quantum couplings.

Typically, the role of the imaginary component in a propagator lies in its introduction of non-Hermitian dynamics and is typically associated with particle decay or absorption \cite{Anastopoulos_2018}. Such formulas often arise in the context of unstable particles whose self-energy picks up an imaginary component. Specifically, the decay width is determined by the formula \(\Gamma = -2 \times \text{Im} \, \Sigma(p)\), where \(\text{Im} \, \Sigma(p)\) is the imaginary part of the self-energy \cite{Anastopoulos_2018}. In the CQ context, we emphasize that the dynamics of the quantum system alone are non-unitary, so such a propagator is to be expected, even in simple systems.

Since the CQ theory induces a coupling to $R^2$ that is real, we know from the same considerations as in $f(R)$ gravity that we require a real contribution to $R_{\mu \nu}R^{\mu \nu}$ in order for the theory to be renormalized. This can only be done by including further higher order terms in the CQ action. For example, in Equation \eqref{eq: CQBDSimple} we could consider contributions such as $-D_0\sqrt{-g}(\Box \phi - \Lambda - c_1 R - c_2 R_{\mu \nu} R^{\mu \nu} + \dots)^2$. However, this induces a higher order $(R_{\mu \nu} R^{\mu \nu})^2$ term in the action. As a consequence, we expect theory introduced in Section \ref{sec: QGScalar} should be viewed as an effective theory. In this way, we also expect that the CQ theory can be stable, even though it includes higher derivative terms in the action, as they are to be treated as perturbations in an EFT. We now discuss further how we can view CQ theory as an effective theory.

\subsection{The CQ theory as an effective theory}\label{sec: cqeffec}

The idea underpinning effective field theory is that one starts by identifying the low energy degrees of freedom that are relevant. One then considers the action which is consistent with the symmetries and physics of the low energy description and writes it in terms of an energy expansion. The terms higher order in the energy expansion are viewed as arising from integrating out unknown degrees of freedom whose propagation is suppressed by the energy scale \cite{GeorgiEFT}. This is the so called bottom up approach, and it enables one to study low energy physics without knowledge of the underlying high energy theory. Such actions are often found approximately from a top down approach. In a top down approach, one starts from a known high energy theory and integrates out high energy modes to study the effective description of the low energy physics \cite{GeorgiEFT}. 

For general relativity, the action in terms of all possible terms that are consistent with gauge symmetry takes the form
\begin{equation}\label{eq: effectiveE}
    S_{\text{eff}} = \int d^4x \sqrt{-g} \left( \Lambda +c_0R + c_1 R^2 + c_2 R_{\mu\nu} R^{\mu\nu} + \ldots \right),
\end{equation}
where \(\Lambda\), \(c_0\), \(c_1\), \(c_2\) are constants to be found experimentally, and the ellipses denote higher powers of \(R\), \(R_{\mu\nu}\) and \(R_{\mu\nu\alpha\beta}\). Actions with higher order curvature terms are known to arise as effective actions in string theory \cite{Green:1987sp}. 

The treatment of quantum gravity as an effective theory is outlined in \cite{Donoghue_1994,wallace2021quantum, Veltman:1975vx, Burgess:2003jk}. The idea is that we can view the gravitational action as being organized in an energy expansion which arises from taking the low energy limit of a UV complete quantum gravity theory. In momentum space, derivatives turn into factors of the four momentum $p$ through \(\nabla_\mu \sim p_\mu\). Hence, each curvature term in Equation \eqref{eq: effectiveE} accumulates a factor of
\(p^2\). Higher order terms in Equation \eqref{eq: effectiveE} therefore involve ever increasing powers of $p^2$. At low energies, terms of order \(p^{2k}\) with $k>1$ are small compared to those of order \(p^2\). This means that the higher order Lagrangian terms will have little effect at low energies compared to the Einstein-Hilbert term. This is one of the reasons that the experimental constraints on higher derivative terms are so small \cite{Donoghue_1994}. 

An important point is that the one loop corrections also involve combinations of curvature terms that are higher order \cite{Donoghue_1994}. Viewed in this way, one can cancel divergences arising from one-loops in the $O(p^2)$ terms by tree-level counter terms involving $O(p^4)$. In this way, one is able to get predictions for the theory valid up to $O(p^4)$ in the energy expansion \cite{Donoghue_1994}. One can then continue the process, calculating observables to within a fixed accuracy dependent on the number of terms included in the action. As one increases the number of terms in the action, the new constants appearing in the action must be fixed by experiment.

The effective theory viewpoint is quite different to that when taking Equation \eqref{eq: quadgravmain} as a UV complete theory \cite{Steinwachs_2020}. In quadratic gravity, one does not consider a perturbation around the free theory defined by the Einstein-Hilbert action, but instead considers $R^2, R_{\mu \nu}R^{\mu \nu} $ in the propagator, which lead to extra degrees of freedom. When considering the loop corrections with these extra degrees of freedom, the divergences cancel amongst themselves. However, when viewed as an effective theory, one instead considers a perturbation around the Einstein-Hilbert action through an infinite series of terms in the action according to Equation \eqref{eq: effectiveE} which leads to different physics.

It would be interesting to study the effective CQ theory of Equation \eqref{eq: fullWF} in more detail, but this is beyond the scope of the current work. This would have to be done from a bottom up approach, since we do not know what the UV completion of quantum gravity, or CQ quantum gravity with a classical scalar field looks like.  Since effective CQ theories have not been studied in the literature, instead, we now argue that one can deal with effective CQ theories in an analogous way to standard quantum field theories: one first specifies a completely positive and normalized low energy effective CQ theory and then proceeds to add all terms consistent with the symmetry of the problem in a power series in the relevant energy scale. We motivate this procedure by studying a top down CQ effective field theory, where the underlying theory is CQ. In the top down case, we have more control over the dynamics, and we can ensure the original dynamics takes the form of a CQ path integral. How CQ theories themselves arise from limits of quantum theories under measurement has been discussed in \cite{Layton:2023wdo}.

In Appendix \ref{sec: effective}, we consider an example where have a normalized completely positive CQ path integral for quantum particles $\psi, \chi$ and a classical variable $q$. We then integrate out the $\chi$ variable to study the effective dynamics of the $\psi, q$ system. 

We start with the CQ action $I = I_{CQ}^+ + I_{CQ}^-$, where 
\begin{equation}\label{eq: effective0main}
\begin{split}
     I_{CQ}[ q, \psi, \chi] & = \int dt \frac{i}{2}\dot{ \psi}^2 + \frac{i}{2}\dot{ \chi}^2 -\frac{i}{2}M^2 \chi^2 - \frac{i g}{4}\chi^2 \psi^2\\
    & -D_0( \ddot{q} - \lambda_{\chi}\chi q - \lambda_\psi \psi q)^2.
\end{split}
\end{equation}
In Equation \eqref{eq: effective0}, $M$ is the mass of the $\chi$ field, $g$ denotes the coupling between $\psi$ and $\chi$, $\lambda_{\chi}$ denotes the coupling between $q$ and $\chi$, and $\lambda_{\psi}$ denotes the coupling between $\psi$ and $q$. We have allowed the $\chi$ field to couple to both $\psi$ and $q$ to illustrate what an effective theory would look like when the high energy field has a coupling to both the classical and quantum degrees of freedom. For simplicity, we take $\psi$ to be without mass.

Expanding out the CQ coupling in Equation \eqref{eq: effective0main}, we can re-write the action as
\begin{equation}
\begin{split}
     I_{CQ}[ q, \psi, \chi] & = \int dt \frac{i}{2}\dot{ \psi}^2 -D_0( \ddot{q} -  \lambda_\psi \psi q)^2\\
     & + \frac{i}{2} \chi( -\frac{d^2}{dt^2} - M^2  -\frac{g}{4} \psi^2 + i D_0 \lambda_{\chi}^2 q^2) \chi\\
     &  +2 D_0 \chi \lambda_{\chi} q(\ddot{q} - \lambda_{\psi} q\psi ).
\end{split}
\end{equation}
We can therefore perform a standard Gaussian integral over the $\chi$ field to find the effective action. In Appendix \ref{sec: effective} we do this explicitly. The final result is that the effective action takes the schematic form 

\begin{equation}\label{eq: integratingoutmain}
\begin{split}
     I_{CQ}^{eff} [ q, \psi] & =\int dt \frac{i}{2}\dot{ \psi}^2 -D_0( \ddot{q} -  \lambda_\psi \psi q)^2  +O(\frac{1}{M}),
\end{split}
\end{equation}
where the $\frac{1}{M}$ expansion involves an infinite series of higher derivative terms arising from integrating out $\chi$. For example, we find a $\frac{1}{M^5}$ term that includes higher derivatives of both the $q$ and $\psi$ fields 
\begin{equation}
    \begin{split}
      &   \frac{\beta}{2 M^5}\big[ ( -\frac{g}{4} \psi^2 + i D_0 \lambda_{\chi}^2q^2)\\
   & \times (-\frac{g}{2} \dot{\psi}^2  -\frac{g}{2} \psi \ddot{\psi} + i2  D_0 \lambda_{\chi}^2\dot{q}^2  +  i 2 D_0 \lambda_{\chi}^2 q \ddot{q}^2) \big],
    \end{split}
\end{equation}
where $\beta$ is a constant that arises from integrating out $\chi$.

Equation \eqref{eq: integratingoutmain} has a leading order term that looks like a standard CQ path integral. Complete positivity and normalization of the dynamics are manifest for this term. This arises due to the direct coupling of $\psi$ and $q$. The remaining terms arises through the effective coupling of $\psi$ and $q$ that is mediated by $\chi$. When calculating the exact effective action, these terms are non-local. However, because the propagation of $\chi$ is suppressed by its mass, we find that for energies less than $M$ the induced interaction terms are approximately local; terms higher order in $\frac{1}{M}$ are generally associated with an increasing number of derivatives \cite{GeorgiEFT}. Each of these higher order terms has its own set of couplings that are to be determined by experiment, but we see that they are related through the underlying CQ theory in Equation \eqref{eq: effective0main}.

Though the final dynamics in Equation \eqref{eq: integratingoutmain} is manifestly completely positive due to the separation of bra and ket fields, it is far from obvious that the dynamics is normalized; the effective dynamics of the $\chi$ field induces unpredictable dynamics in the effective CQ action. This is true even though the action we started with in Equation \eqref{eq: effective0main} defines a valid CQ path integral, and so we know the total dynamics is normalized. The same phenomena occurs when studying the effective actions of unitary QFT's. In general, the effective action consists of a leading order term describing a unitary theory with the remaining terms being higher order, arranged in a power expansion determined by the energy scales of the original action. If one truncates the series, one finds the effective dynamics is no-longer unitary \cite{GeorgiEFT}. More generally, extrapolating the physics of the first few terms to high energies will lead to a different description than that of the full theory. It is in this way that the differences arise when treating quadratic gravity as a fundamental theory versus an effective one.

Because of the similarities between the top down CQ theory and the standard effective field theories, we suggest that we construct effective CQ theories in a similar way. Indeed, the CQ path integral can be viewed in the language of a quantum field theory, and the arguments underpinning effective field theory are general \cite{GeorgiEFT}. In constructing an effective CQ theory, one first specifies a low energy CQ action describing the relevant physics, such as that in Equation \eqref{eq: fullWF}. One then then specifies, as a power series in an energy scale, all of the allowed CQ couplings by symmetry which are to be fixed via experiment. The higher order terms, up to the energy scale of interest, should have couplings that are fixed by experiment. Truncating at finite $M$ then gives dynamics that is not completely positive normalized CQ dynamics, but this is to be expected, since in this case we are ignoring terms that are required for the overall dynamics to be CP and norm preserving. In this way it is possible to study an effective CQ theory without knowing the underlying CQ theory that generates it. It would be interesting to try and perform this procedure for the peturbative CQ theory of Equation \eqref{eq: fullWF}. In particular to see how one computes precisely low energy observables such as the decoherence rate.

\section{Discussion}\label{sec: discussion} 
In this work, we have studied a theory of quantum gravity coupled to a classical scalar field with a diffeomorphism invariant action. The coupling to the scalar field decoheres the quantum system according to the Ricci scalar of the quantum system, while the classical limit predicts diffusion of the effective Newtonian constant. 

The theory provides another example of diffeomorphism invariant classical-quantum dynamics \cite{Oppenheim:2023izn, Oppenheim:2024iae} and, to our knowledge, the first example of a diffeomorphism invariant quantum gravity collapse model. In this work, the loss of coherence is a derived consequence of the interaction of a quantum spacetime with a classical scalar field. The theory is not renormalizable, but should be viewed as an effective theory. Nonetheless, we find it an interesting effective theory since a measurement postulate is not necessarily required: if one is to consider quantum measurement as a low energy phenomena that should be describable by dynamics, then classical scalar fields coupled to quantum gravity are potential candidates. 

Though the theory is to be understood as effective, this does not preclude the scalar field from being fundamentally classical, and the theory could be considered as a low energy limit of a fundamental CQ theory. Alternatively, the scalar field could be effectively classical as part of a limit of a fundamental quantum theory \cite{Layton:2023wdo}. It would be interesting to try to integrate the dynamics of classical scalar fields with models of UV complete quantum gravity.

Since CQ path integrals need to be better understood, it is important to explore the dynamics in more detail, and our work leads to many open questions: How should we view an effective CQ theory? Are there clues as to what sort of high energy theories yield such a theory in the low energy limit? What do such theories predict on cosmological scales? How does one go about renormalizing a fundamental CQ theory? We leave these as open questions for future work.

Though we view the CQ gravity theory as an effective theory, it would be interesting to understand whether or not CQ theories can be renormalized. This would be important if they are to be treated as fundamental. In particular, there may be subtleties when trying to preserve the complete positivity of the combined CQ interaction under renormalization. Though the CQ path integrals do not couple the bra and ket fields directly, it has been shown that power counting arguments can often fail in open quantum field theory. In particular,  \cite{Baidya:2017eho} show that an open   $\phi^3 + \phi^4$ theory is renormalizable, but \cite{Avinash:2019qga} have showed that open Yukawa theory is not.

We chose the Brans-Dicke coupling because it was the most computationally tractable, but there are many general conclusions to be learned from the theory. For example, we expect that non-minimally coupled actions containing a scalar field and metric that have second-order equations of motion will generically have similar properties to the Brans-Dicke theory. In particular, if the equations of motion for the scalar field depend on the Ricci scalar, as is fairly generic in Horndeski theories \cite{Horndeski1974, Kobayashi_2019}, then this will also provide a mechanism of decoherence, where superpositions of quantum spacetimes with differing Ricci tensor decohere. Studying the cosmological effects of CQ Horndeski theories would be interesting. Scalar couplings to gravity also arise naturally in string theory \cite{Green:1987sp, Easson:2020bgk}, and it is curious if one can arrive at the CQ theory considered here by taking a classical-quantum limit of a quantum gravity system \cite{Layton:2023wdo}. For example, to see if one can arrive at a classical field coupled to gravity from considerations of decoherence in quantum cosmology \cite{Colas:2024xjy,Halliwell:1989vw,Dowker:1992es,Kiefer:1998qe,Colas:2023wxa}.

Another aspect of quantum gravity that needs to be touched upon is the constraints of the gravitational theory. These will be altered due to the coupling with the scalar field. We have considered a theory based on configuration space variables to make the spacetime symmetries manifest. However, the constraints are most easily seen in the phase space representation. Constraints for the CQ theories of classical gravity were studied in \cite{pqconstraints}, while the Newtonian limit of CQ gravity was studied in detail in \cite{layton2023weak} building on work by \cite{2016Tilloy}. It was shown that they are stochastic in nature. It is important to explore how the constraint structure is altered in the present theory when gravity is kept quantum.

\section*{Acknowledgements}
We would like to thank Maite Arcos, Isaac Layton, Jonathan Oppenheim, Emanuele Panella, Andrea Russo and Andy Svesko for valuable discussions. We thank Jules Tilly for reading through a draft version of the manuscript.
\bibliography{refCQ}

\appendix

\section{Ensuring the CQ path integral is normalized}\label{section: normalizing}
In this section, we show that the CQ action defined by Equation's \eqref{eq: normalizedCQmain} and \eqref{eq: normalizedCQmainLower} are normalized.

To see the problem of normalization of higher derivative path integrals in more detail, we will review how the normalization of quantum states occurs in Lindbladian path integrals with a Feynman-Vernon action \cite{FeynmanVernon}, and how probabilities are conserved in higher-order classical path integrals. Let us first consider higher-order classical path integrals. We refer the reader to \cite{UCLPILONG} for a complete derivation of normalized CQ path integrals from master equations. 

\subsection{Normalization of higher derivative classical path integrals}
When considering a classical path integral that contains higher derivatives, we should treat $q, \dot{q}$ as independent variables. This is outlined in detail in \cite{Hawking_2002}. To that end, we will show how the normalization of the path integral
\begin{equation}\label{eq: classPI}
    \p(q_f, \dot{q}_f,t_f) = \int^{B} \ \mathcal{D}q e^{-\int_{t_i}^{t_f} dt[\ddot{q}- f(\dot{q}, q)  ]^2}p(q_i, t_i)
\end{equation}
occurs. In Equation \eqref{eq: classPI}, note that the boundary conditions are given by $B=\{q(t_f) = q_f, \dot{q}(t_f) = \dot{q}_f\}$, which involve both  $q$ and  $\dot{q}$.

To check normalization, we consider Equation \eqref{eq: classPI} for small $\delta t$, with $t_n = \delta t + t_{n-1}$
\begin{equation}\label{eq: shorttimeclass}
\begin{split}
   &  \p(q_{n+1}, q_{n+2},t_{n+1}) =  \int dq_{n} e^{-\delta t[\frac{q_{n+2}-2q_{n} + q_{n+1}}{\delta t}-f(q_{n+1}, q_n) ]^2}\\
   & \times p(q_n, q_{n+1}, t_n).
    \end{split}
\end{equation}
The norm of the probability distribution is found by performing the integral over the final variables $q_{n+1}, q_{n+2}$
\begin{equation}\label{eq: highernorm}
  \begin{split}
 & \int dq_{n} dq_{n+1} dq_{n+1}  e^{-\delta t[\frac{q_{n+2}-2q_{n} + q_{n+1}}{\delta t}-f(q_{n+1}, q_n) ]^2}\\
 & \times p(q_n, q_{n+1}, t_n).
\end{split}  
\end{equation}
Equation \eqref{eq: highernorm} defines a standard Gaussian integral over the $q_{n+2}$ coordinate. Hence, the $q_{n+2}$ integral eats the action up to a Gaussian normalization factor that we can calculate exactly, and we are left with 
\begin{equation}
1=  \int dq_{n} dq_{n+1} N p(q_n, q_{n+1}, t_n),
\end{equation}
so we can simply absorb $N$ into the measure, and the path integral will be normalized. If we were to include a $q$ dependent diffusion coefficient $D_2(q,\dot{q})$ in Equation \eqref{eq: classPI}, then the Gaussian integral will be $q$ dependent, and this will need to be included in the measure for $\mathcal{D}q$. One can also re-exponentiate this term via a Faddeev-Popov action, as in Equation \eqref{eq: Fadeev}. The message is that the higher derivative terms in the classical path integral are standard Gaussian integrals if we consider $q$ and $\dot{q}$ as independent variables. Hence, the classical contribution to Equation \eqref{eq: CQBD}, which also involves the $\pm$ bra-ket branch average, will be normalized in a similar way. We show this explicitly in Section \ref{sec: CQnormalizationApp} 

\subsection{Normalization of higher derivative Feynman-Vernon path integrals}
Let us now consider a Feynman-Vernon quantum path integral with a decoherence term. Consider first the path integral for a quantum state $\sigma$
\begin{equation}\label{eq: lind0}
\begin{split}
     & \sigma(\psi^+_f, \psi^-_f, t_f) = \int^B \mathcal{D} \psi^+ \mathcal{D}\psi^-  \\
     & e^{\int_{t_i}^{t_f}dt i[\dot{\psi_+}^{2} + V(\psi_+)] -i[\dot{\psi_-}^{2} + V(\psi_-)]  -\frac{D_0}{2}(L(\psi_+)-L(\psi_-))^2}\\
     & \times \sigma(\psi_i^+, \psi^-_i, t_i),
    \end{split}
\end{equation}
where $B$ imposes the final state boundary conditions on the bra and ket fields, and $L(\psi)$ is an arbitrary operator of $\psi$ but not of its derivatives. 

For Equation \eqref{eq: lind0}, it will prove insightful to show how the kinetic term enforces the normalization of the quantum state. To that end, consider the short time version of Equation \eqref{eq: lind}

\begin{equation}\label{eq: lind}
\begin{split}
     & \sigma(\psi^+_{n+1}, \psi^-_{n+1}, t_f)  = \int d\psi^+_n d\psi^-_n \\
     & \times e^{\delta t [i(\frac{\psi^+_{n+1} -\psi^+_n }{\delta t})^{2} + iV(\psi^+_n) -i(\frac{\psi^-_{n+1} -\psi^-_n }{\delta t})^{2} -i V(\psi^-_n)] }\\
    & e^{ -\delta t \frac{D_0}{2}(L(\psi_n^+)-L(\psi_n^-))^2}\sigma(\psi_n^+, \psi^-_n, t_n).
    \end{split}
\end{equation}
The trace of the quantum state is found by matching the $\psi^+_{n+1}= \psi^-_{n+1} = \psi$ fields and integrating over $\psi$
\begin{equation}
\begin{split}
    &  \int d\psi^+_n d\psi^-_n d\psi  \times e^{\frac{i}{\delta t} \psi( \psi_n^+- \psi_n^-)} e^{i \delta t [V(\psi_n^+) -i V(\psi_n^-)] }\\
    & e^{ -\delta t \frac{D_0}{2}(L(\psi_n^+)-L(\psi_n^-))^2}\sigma(\psi_n^+, \psi^-_n, t_n).
    \end{split}
\end{equation}
Performing the integration over $\psi$ gives rise to a delta function $\delta(\psi^+_n -\psi^-_n)$. Hence, the quantum state is normalized to constant factors that can be absorbed.

However, had we included higher-order kinetic terms in the decoherence sector, we would not have found this normalization. In particular, if the decoherence term was instead
\begin{equation}
\begin{split}
   \int_{t_i}^{t_f}dt \frac{1}{2}D_0( \dot{\psi}_+^2 - \dot{\psi}_-^2)^2,
    \end{split}
\end{equation}
then the delta function integral is not imposed, and the state is no longer normalized to constant factors.

As such, it seems inevitable that for the path integral to be normalized with higher derivative decoherence terms, one needs to also add higher derivative kinetic terms in the action. In this case, the action 
\begin{equation}
\begin{split}
    S &= \int_{t_i}^{t_f}dt i[\dot{\psi_+}^{2} + \ddot{\psi_+}^{2} - V(\psi_-)] -i[\dot{\psi_-}^{2} + \ddot{\psi_-}^{2} - V(\psi_-)]\\
    & - \frac{1}{2}D_0( \dot{\psi}_+^2 - \dot{\psi}_-^2)^2
    \end{split}
\end{equation}
is normalized up to constant factors by the same argument, so long as we treat $\psi $ and $\dot{\psi}$ as independent variables to be specified in the quantum state; this is also argued for independent reasons in \cite{Hawking_2002}. To see this, one does the short time expansion, treating $\psi $ and $\dot{\psi}$ as independent variables as in the higher derivative classical path integral. Computing the trace then sets $\psi_{n+2}^{\pm}$ equal to each other, as well as the setting the $\psi_{n+1}^{\pm}$ fields equal to be the same. The $\psi_{n+2}$ integral then enforces a delta function over $\delta(\psi^+_n-\psi^-_n)$, which kills the decoherence term and means that the path integral is normalized up to constant factors. 
\subsection{Normalization of CQ path integrals}\label{sec: CQnormalizationApp}
In this section, we show that Equation's \eqref{eq: normalizedCQmainLower} give rise to normalized CQ dynamics \eqref{eq: normalizedCQmain}. The proofs will follow in the same way as the discussion of normalization of classical and quantum path integrals.

Let us start with the higher derivative case. We show that any CQ path integral with action 
\begin{equation}\label{eq: normalizedCQApp}
\begin{split}
    & I[q,\psi^+, \psi^-]= \int dt i \ddot{\psi}_+^2 + i V(\psi^+, \dot{\psi}^+) - i \ddot{\psi}_-^2 -i V(\psi^-, \dot{\psi}^-) \\
    &- \frac{D_0(q,\dot{q}, \psi^+, \dot{\psi^+})}{2}(\ddot{q} + f(q,\dot{q}, \psi^+, \dot{\psi}^+))^2 \\
    & - \frac{D_0(q,\dot{q}, \psi^-, \dot{\psi^-})}{2}(\ddot{q} + f(q,\dot{q}, \psi^-, \dot{\psi}^-))^2
    \end{split}
\end{equation}
is normalized up-to constant factors when $D_0>0$. In the case where $D_0$ has a functional dependence on the fields one must make sure to also include a factor of  $\sqrt{\det(D_0(q,\dot{q}, \psi)}$ in the path integral measure. Equation \eqref{eq: normalizedCQApp} is a generic type of action one gets from varying Equation \eqref{eq: CQProto2} with a CQ proto action that has second order equations of motion for the classical degree of freedom. 

The steps in showing Equation \eqref{eq: normalizedCQApp} follow in the same way as the discussions of classical and quantum path integrals. Firstly, because the action is higher derivative, the CQ state is specified through $\cqstate(q,\dot{q}, \psi^{\pm}, \dot{\psi}^{\pm})$. 

Taking the trace at the $t_{n+1} = t_{n}+ \delta t$ time-step therefore involves identifying $\psi_{n+2}^+=\psi_{n+2}^- = \psi_{n+2}$ and $\psi_{n+1}^+=\psi_{n+1}^-=\psi_{n+1}$. We then integrate over the $\psi_{n+2}$ and $\psi_{n+1}$ variables, as well as over the $q_{n+2}, q_{n+1}$ classical degrees of freedom.

Let us first look at the higher derivative quantum kinetic term. This can be expanded as
\begin{equation}
\begin{split}
   &  \ddot{\psi}_+^2 -\ddot{\psi}_-^2 \sim \\
   & (\psi_{n+2} - 2\psi_n^+ + \psi_{n+1})^2 - (\psi_{n+2} - 2\psi_n^- + \psi_{n+1})^2\\
   & = 4\big[ (\psi_{n}^+)^2-(\psi_{n}^-)^2 + \psi_{n+2}( \psi_n^--\psi_n^+) + \psi_{n+1}(\psi_n^- -\psi_n^+) \big].
    \end{split}
\end{equation}
Hence, integrating over $\psi_{n+2}$ gives a delta function in $\delta(\psi_n^--\psi_n^+)$. As a consequence of this, all the bra and ket fields in the path integral are identified. We are therefore left with the action 
\begin{equation}\label{eq: gaussian_final_cq}
\begin{split}
    & I'[q,\psi ]= -\int dt D_0(q,\dot{q}, \psi,\dot{\psi})(\ddot{q} + f(q,\dot{q}, \psi, \dot{\psi}))^2 .
    \end{split}
\end{equation}
Since all the bra and ket quantum fields are identified, normalization of Equation \eqref{eq: normalizedCQApp} is equivalent to ensuring that Equation \eqref{eq: gaussian_final_cq} is normalized.

As we saw for the classical path integrals, integrating Equation \eqref{eq: gaussian_final_cq} over the $\ddot{q}$ at second time step implements a standard Gaussian integral. If $D_0$ is dependent on the fields, we therefore pick up a term $(\sqrt{\det(D_0(q,\dot{q}, \psi)})^{-1/2},
$
which we must cancel in the measure by including a $\sqrt{\det(D_0(q,\dot{q}, \psi)}$ term. It can also be exponentiated into the action by introducing Bosonic and Fermionic Faddeev-Poppov fields \cite{Bastianelli_2017}. This term commonly arises in the study of Fokker-Plank type equations when the noise is multiplicative \cite{Onsager1953Fluctuations,Dekker, Bastianelli_2017}. With this in mind,  once we have integrated over $\ddot{q}$, the action vanishes and we are left with the normalization of the initial CQ state. Hence the path integral preserves the normalization of CQ states.

For the present work, we have a non-minimal coupling of the classical scalar field to gravity, and so we must include higher derivative terms in the gravitational sector of the CQ action, as in Equation \eqref{eq: BDNorm}.

In a similar manner, we can also show that the path integral of Equation \eqref{eq: normalizedCQmainLower} is also normalized 
\begin{equation}\label{eq: normalizedCQmainLowerAppendix}
\begin{split}
    & I[q,\psi^+, \psi^-]= \int dt i \dot{\psi}_+^2 + i V(\psi^+) - i \dot{\psi}_-^2 -i V(\psi^-) \\
    &- \frac{D_0(q,\dot{q}, \psi^+)}{2}(\ddot{q} + f(q,\dot{q}, \psi^+))^2 \\
    & - \frac{D_0(q,\dot{q}, \psi^-)}{2}(\ddot{q} + f(q,\dot{q}, \psi^-))^2,
    \end{split}
\end{equation}
where $D_0>0$. To see this, we first take the trace of the system, setting $\psi_{n+1}^+= \psi_{n+1}^- =\psi$. Integrating over $\psi$ then enforces a delta function $\delta( \psi^+-\psi^-)$. We are then left with the action 
\begin{equation}\label{eq: gaussian_final_cq_lower}
\begin{split}
    & I'[q,\psi ]= -\int dt D_0(q,\dot{q}, \psi)(\ddot{q} + f(q,\dot{q}, \psi))^2, 
    \end{split}
\end{equation}
and we can again perform the Gaussian integral over $\ddot{q}$ to arrive at a normalized path integral if $\sqrt{\det(D_0(q,\dot{q}, \psi)}$ is included in the measure.

\section{Effective CQ dynamics}\label{sec: effective}

In this section we consider an effective CQ theory via a top down approach. We study an example where have a CQ path integral for quantum particles $\psi, \chi$ and a classical variable $q$. We then integrate out the $\chi$ variable to study the effective dynamics of the $\psi, q$ system. This section follows in a similar way to the derivation of quantum effective actions given in \cite{Skinner2023AQFT}.

We start with the CQ action $I = I_{CQ}^+ + I_{CQ}^-$, where 
\begin{equation}\label{eq: effective0}
\begin{split}
     I_{CQ}[ q, \psi, \chi] & = \int dt \frac{i}{2}\dot{ \psi}^2 + \frac{i}{2}\dot{ \chi}^2 -\frac{i}{2}M^2 \chi^2 - \frac{i g}{4}\chi^2 \psi^2\\
    & -D_0( \ddot{q} - \lambda_{\chi}\chi q - \lambda_\psi \psi q)^2.
\end{split}
\end{equation}
In Equation \eqref{eq: effective0}, $M$ is the mass of the $\chi$ field, $g$ denotes the coupling between $\psi$ and $\chi$, $\lambda_{\chi}$ denotes the coupling between $q$ and $\chi$, and $\lambda_{\psi}$ denotes the coupling between $\psi$ and $q$. We have allowed the $\chi$ field to couple to both $\psi$ and $q$ to illustrate what an effective theory would look like if the unknown field has a universal coupling to both the classical and quantum degrees of freedom. 
Expanding out the CQ coupling in Equation \eqref{eq: effective0}, we find we can re-write the action as
\begin{equation}
\begin{split}
     I_{CQ}[ q, \psi, \chi] & = \int dt \frac{i}{2}\dot{ \psi}^2 -D_0( \ddot{q} -  \lambda_\psi \psi q)^2\\
     & + \frac{i}{2} \chi( -\frac{d^2}{dt^2} - M^2  -\frac{g}{4} \psi^2 + i D_0 \lambda_{\chi}^2 q^2) \chi\\
     &  +2 D_0 \chi \lambda_{\chi} q(\ddot{q} - \lambda_{\psi} q\psi ).
\end{split}
\end{equation}
We can therefore perform a standard Gaussian integral over the $\chi$ field to find the effective action. Defining
\begin{equation}
    A(t)= -\frac{d^2}{dt^2} - M^2 -\frac{g}{4} \psi^2 + i D_0 \lambda_{\chi}^2 q^2 ,
\end{equation}
we find the effective action is given by
\begin{equation}\label{eq: integratingout}
\begin{split}
     I_{CQ}^{eff} [ q, \psi] & =\int dt \frac{i}{2}\dot{ \psi}^2 -D_0( \ddot{q} -  \lambda_\psi \psi q)^2\\
     &  +\frac{1}{2}A^{-1} i (2 D_0 \lambda_{\chi} q(\ddot{q} - \lambda_{\psi} \psi  q)^2)^2\\
     & -\frac{1}{2} \ln \det A.
\end{split}
\end{equation}
In Equation \eqref{eq: integratingout} we find that we have a standard looking CQ dynamics involving the variables $q, \psi$, but also extra interactions that arise through the integrated out $\chi$ field. We now explore these extra contributions. We first focus on the $\ln \det A$ term. This can be simplified using the formula 
\begin{equation}\label{eq: lndet}
\begin{split}
 \ln \det A& = \Tr \ln(B) \\
 & + \Tr \ln( 1-B^{-1}(-\frac{g}{4} \psi^2 + i D_0 \lambda_{\chi}^2 q^2)).
\end{split}
\end{equation}
where we have defined
\begin{equation}
  B=-\frac{d^2}{dt^2} - M^2. 
\end{equation}
The first term in Equation \eqref{eq: lndet} is independent of the fields, so just contributes an absorbable constant factor. 
 $B^{-1}= G(t,t')$ is the solution to the Greens function 
\begin{equation}
    (-\frac{d^2}{dt^2} - M^2 )G(t,t') = \delta(t-t'),
\end{equation}
and has the solution $G(t,t') = \frac{i}{2M}e^{i M|t-t'|}$.

Using the Greens function, we can expand the $\ln \det A$ term in Equation \eqref{eq: integratingout} to find 
\begin{equation}\label{eq: nonlocal}
\begin{split}
   & \tr \ln( 1-B^{-1}(-\frac{g}{4} \psi^2 + i D_0 \lambda_{\chi}^2)q^2) =\\
   & - \int dt G(t,t)( -\frac{g}{4} \psi^2(t) + i D_0 \lambda_{\chi}^2)q^2(t)) \\
   & -\frac{1}{2} \int dt dt' G(t,t') ( -\frac{g}{4} \psi^2(t) + i D_0 \lambda_{\chi}^2q^2(t))G(t,t')\\
   & \times ( -\frac{g}{4} \psi^2(t') + i D_0 \lambda_{\chi}^2q^2(t')) + \dots
\end{split}
\end{equation}
where the $\dots$ denotes the higher order non-local terms that arise from expanding the logarithm. Equation \eqref{eq: nonlocal} is manifestly non-local. However, for $M|t-t'|$ large, the oscillations from the positive and negative parts in the integrand will cancel out. Hence, we can expand the fields $\psi(t'), q(t')$ around $t$ to find an approximate form of Equation \eqref{eq: nonlocal}. For example, the term in Equation \eqref{eq: nonlocal} involving the double time integral now takes the form
\begin{equation}\label{eq: nonlocal2}
\begin{split}
   & -\frac{1}{2} \int dt dt' G(t,t')^2 ( -\frac{g}{4} \psi^2(t) + i D_0 \lambda_{\chi}^2q^2(t))\\
   & \times \big( -\frac{g}{4} \psi^2(t) + i D_0 \lambda_{\chi}^2q^2(t) \\
   &   +[- \frac{g}{2} \psi \dot{\psi}(t) + i 2 D_0 \lambda_{\chi}^2q \dot{q}(t))](t'-t)\\
   & +  [-\frac{g}{2} \dot{\psi}^2(t)  -\frac{g}{2} \psi \ddot{\psi}(t) + i 2 D_0 \lambda_{\chi}^2\dot{q}^2 (t) +  i 2 D_0 \lambda_{\chi}^2 q \ddot{q}^2(t)](t'-t)^2\\
   & + \dots \big),
\end{split}
\end{equation}
where the $\dots$ denote higher order terms in $(t-t')$ that we have neglected. These involve increasing derivatives of the fields. 

In Equation \eqref{eq: nonlocal2} we can now perform the $t'$ integral \cite{Skinner2023AQFT}. First, we note that terms odd in $(t-t')$ will not contribute, since $G(t,t')^2 \sim \frac{-1}{ M^2} e^{2iM|t-t'|}$. The remaining terms are found by noting that $G(t,t')^2$ involves a factor of $\frac{1}{M^2}$ and depends only on $t'$ through the combination $u=M(t-t')$. Hence, replacing the $(t'-t)^n$ terms with $(u/M)^n$ and changing variables $dt'=\frac{du}{M}$ then we find the integral over $u$ will just yield some prefactor constants. We find that Equation \eqref{eq: nonlocal2} then takes the form of a local integral 
\begin{equation}\label{eq: localEff}
\begin{split}
   & \frac{1}{2} \int dt \frac{\alpha}{M^3}( -\frac{g}{4} \psi^2(t) + i D_0 \lambda_{\chi}^2q^2(t))^2 \\
   & + \frac{\beta}{M^5}\big[ ( -\frac{g}{4} \psi^2(t) + i D_0 \lambda_{\chi}^2q^2)\\
   & \times (-\frac{g}{2} \dot{\psi}^2  -\frac{g}{2} \psi \ddot{\psi} + i2  D_0 \lambda_{\chi}^2\dot{q}^2  +  i 2 D_0 \lambda_{\chi}^2 q \ddot{q}^2) \big] \\
   & + \frac{1}{M^7}(\text{four-derivative terms})+ \dots. 
\end{split}
\end{equation}
In Equation \eqref{eq: localEff} $\alpha$ and $\beta$ are numbers found from the integration over $t'$. 

Similarly, the factor of $A$ appearing in Equation \eqref{eq: integratingout} looks like a propagator with a $\psi$ dependent mass. It also has a decaying component because of the imaginary $D_0$ contribution. Since $G(t,t)\sim \frac{1}{M}$, to leading order in $M$, we expect that we can write
\begin{equation}
     +\frac{1}{2}A^{-1} i (2 D_0 \lambda_{\chi} q(\ddot{q} - \lambda_{\psi} \psi  q)^2)^2 \approx \frac{i \gamma}{M}  (q(\ddot{q} - \lambda_{\psi} \psi  q)^2)^2,
\end{equation}
for some constant $\gamma$. There will also be higher order terms that we do not know how to calculate but that we expect can be represented by a power expansion in $\frac{1}{M}$. 

In total, we find the effective action takes the schematic form 
\begin{equation}\label{eq: integratingoutApp}
\begin{split}
     I_{CQ}^{eff} [ q, \psi] & =\int dt \frac{i}{2}\dot{ \psi}^2 -D_0( \ddot{q} -  \lambda_\psi \psi q)^2  +O(\frac{1}{M})
\end{split}
\end{equation}
where the $\frac{1}{M}$ expansion involves lots of complex terms arising from integrating out $\chi$ that are hard to predict the structure of.

The main point is that, through an expansion in $\frac{1}{M}$, we expect that we can write the effective CQ action as a local action, but in terms of general higher derivative terms in both $q$ and $\psi$. These higher derivatives are suppressed by factors of $\frac{1}{M}$. Intuitively, the higher derivative terms arise because $\psi$ couples to $q$ indirectly through $\chi$. These couplings are non-local. However, because the propagation of $\chi$ decays due to its mass, they can be expressed in terms of a local $\frac{1}{M}$ power expansion in the low energy regime.  Since this discussion of the effective CQ theory follows in the same way as standard EFTs, we therefore expect that we can treat effective CQ theories on the same footing as standard EFT's \cite{GeorgiEFT}.

\onecolumngrid
\newpage

\end{document}

%% file: definitions.tex
\newcommand{\q}{q}
\newcommand{\p}{p}

\newcommand{\xql}{\psi^-}
\newcommand{\xqr}{\psi^+}

\def\<{\langle}
\def\>{\rangle}

\newcommand{\cqstate}{\varrho}

\newcommand{\be}{\begin{eqnarray} \begin{aligned}}
\newcommand{\ee}{\end{aligned} \end{eqnarray} }
\newcommand{\benn}{\begin{eqnarray*} \begin{aligned}}
\newcommand{\eenn}{\end{aligned} \end{eqnarray*} }

\newcommand{\ben}{\begin{eqnarray} \begin{aligned}}
\newcommand{\een}{\end{aligned} \end{eqnarray} }

\newcommand{\bc}{\begin{center}}
\newcommand{\ec}{\end{center}}

\newcommand{\tr}{\mathop{\mathsf{tr}}\nolimits}
\newcommand{\Tr}{\mathop{\mathsf{Tr}}\nolimits}

\newcommand{\beq}{\begin{eqnarray} \begin{aligned}}
\newcommand{\eeq}{\end{aligned} \end{eqnarray} }
\newcommand{\bea}{\begin{array}}
\newcommand{\eea}{\end{array}}

\newcommand{\bee}{\begin{enumerate}}
\newcommand{\eee}{\end{enumerate}}
\newcommand{\bei}{\begin{itemize}}
\newcommand{\eei}{\end{itemize}}

\usepackage{amsfonts}

\def\01{\{0,1\}}

\def\<{\langle}
\def\>{\rangle}

\newtheorem*{rep@theorem}{\rep@title}
\newcommand{\newreptheorem}[2]{%
\newenvironment{rep#1}[1]{%
 \def\rep@title{#2 \ref{##1} (restatement)}%
 \begin{rep@theorem}}%
 {\end{rep@theorem}}}
\makeatother

\newreptheorem{thm}{Theorem}
\newreptheorem{lem}{Lemma}